\documentclass[11pt]{article}
\usepackage[margin=1in]{geometry}

\usepackage[utf8]{inputenc}

\usepackage[dvipsnames]{xcolor}

\usepackage{amsmath,amssymb}
\usepackage{nameref}
\usepackage{hyperref}
\usepackage{verbatim}

\usepackage[aboveskip=1pt,labelfont=bf,labelsep=period,justification=raggedright,singlelinecheck=off]{caption}
\usepackage{graphicx}

\usepackage{tikz}
\usetikzlibrary{shadows.blur}
\usepackage{enumitem}
\setitemize{noitemsep,topsep=0pt,parsep=0pt,partopsep=0pt}

\usepackage[
    natbib=true,
    style=numeric-comp,
    sorting=none,
]{biblatex}
\newcommand{\MOne}{M\textsubscript{1}}
\newcommand{\MTwo}{M\textsubscript{2}}
\newcommand{\zzIm}{\mathrm{PI}^{\mathrm{zz}}_0(M)}
\newcommand{\vinIm}{\mathrm{PI}^{\mathrm{vin}}_0(M)}
\newcommand{\radIm}{\mathrm{PI}^{\mathrm{rad}}_0(T)}
\newcommand{\RipsIm}{\mathrm{PI}^{\mathrm{VR}}}

\newcommand{\PH}{\operatorname{PH}}

\begin{document}

\title{\bf Topological classification of tumour-immune interactions and dynamics}
\author{
Jingjie Yang,
Heidi Fang,
Jagdeep Dhesi,
Iris H.R. Yoon, \\
Joshua A. Bull,
Helen M. Byrne,
Heather A. Harrington,
Gillian Grindstaff*
\\
\textit{Mathematical Institute, University of Oxford, Oxford, OX2 6GG, UK}\\
\textit{*University of California, Los Angeles, 90095, USA}
}
\date{}
\maketitle

\begin{abstract}
The complex and dynamic crosstalk between tumour and immune cells results in tumours that can exhibit distinct qualitative behaviours---elimination, equilibrium, and escape---and intricate spatial patterns, yet share similar cell configurations in the early stages.
We offer a topological approach to analyse time series of spatial data of cell locations (including tumour cells and macrophages) in order to predict malignant behaviour. 
We propose four topological vectorisations specialised to such cell data: persistence images of Vietoris-Rips and radial filtrations at static time points, and persistence images for zigzag filtrations and persistence vineyards varying in time. 
To demonstrate the approach, synthetic data are generated from an agent-based model with varying parameters. We compare the performance of topological summaries in predicting---with logistic regression at various time steps---whether tumour niches surrounding blood vessels are present at the end of the simulation, as a proxy for metastasis (i.e., tumour escape).
We find that both static and time-dependent methods accurately identify perivascular niche formation, significantly earlier than simpler markers such as the number of tumour cells and the macrophage phenotype ratio. 
We find additionally that dimension 0 persistence applied to macrophage data, representing multi-scale clusters of the spatial arrangement of macrophages, performs best at this classification task at early time steps, prior to full tumour development, and performs even better when time-dependent data are included; in contrast, topological measures capturing the shape of the tumour, such as tortuosity and punctures in the cell arrangement, perform best at intermediate and later stages. We analyse the logistic regression coefficients for each method to identify detailed shape differences between the classes.
\end{abstract}


\section{Introduction}

The ecosystem of cells that surround a tumour— the tumour micro-environment— is a complex system. The interplay of tumour cells with immune cells and the surrounding blood vessels and tissue play a crucial role in the evolution of behaviour of this system. For example, interactions with the tumour can lead the immune cells to either support or inhibit tumour cell proliferation and migration. An overarching goal in mathematical oncology is the study and prediction of the tumour behaviour from earlier observations, which may improve understanding of the complex system as well as inform appropriate treatment strategies. With advances in technologies and more sophisticated models, spatio-temporal data of species in the tumour microenvironment are more readily available. We generate data using agent-based models that encode species interactions in the tumour microenvironment, which provides species locations at multiple different conditions and multiple time points. 
 We propose the mathematical machinery of \emph{persistent homology} to quantify the spatio-temporal evolution of tumour and immune species. We compare the spatio-temporal analysis using topological descriptors with simpler descriptors, such as tumour size, macrophage phenotype ratio and minimum distance between tumour and blood vessel.
Our aim is to analyse, classify and predict different cellular behaviours with direct consequence to tumour outcome. 

%

%

Researchers have long known that innate immune cells such as macrophages can exhibit both pro- and anti-tumour behaviours depending on their phenotype \cite{immune-tumour}. The relationships between the spatial and phenotypic distributions of macrophages and patterns of tumour growth are believed to hold great significance for prognosis \cite{prognostic} and responses to cancer immunotherapy \cite{immunotherapy}.
Many models of macrophage-tumour interactions have been developed (eg, \cite{model1,model2,model3,model4,model5,model6}), including models of the spatial interactions between macrophages and tumour cells required for the development of perivascular niches (such as the CSF-1/EGF paracrine loop driving cross-talk between pro-tumour macrophages and tumour cells \cite{Knutsdottir2014,Knutsdottir2016,Elitas2016}) and models in which macrophage phenotype is resolved as a continuum (eg \cite{Eftimie2020,model6}). Here
 we will restrict analysis to synthetic data generated previously 
 by a two-dimensional, off-lattice 
 \textit{agent-based model} (ABM) by Bull and Byrne \cite{Josh:ABM}, in which a wide range of tumour-macrophage interactions are determined based on a continuum of macrophage phenotypes.
The ABM distinguishes four cell types---macrophages, stromal, tumour, and necrotic cells.
Each cell is represented by an agent whose behaviour is determined by a set of rules: each cell is subject to mechanical and chemotactic forces exerted by neighbouring cells, and the net force determines how its spatial position changes over time.
%
\begin{figure}[h]
\includegraphics[width=\textwidth]{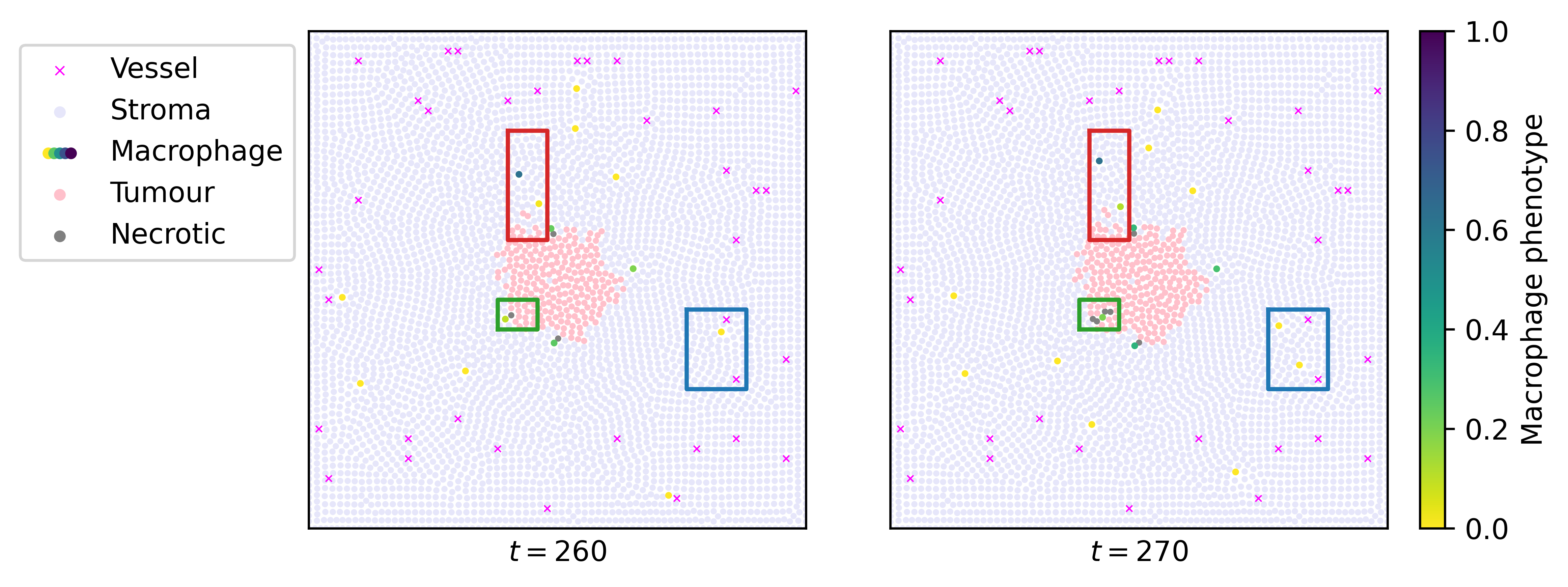}
\caption{{\bf Snapshots of an ABM simulation illustrating how macrophage behaviour
changes with phenotype.} \MOne macrophages have phenotype values closer to 0 (yellow/green), whereas \MTwo macrophages have phenotype values closer 
to 1 (blue).
Blue box: \MOne-like macrophages extravasate from blood vessels and migrate towards the tumour.
Green box: An \MOne-like macrophage infiltrates the tumour and kills tumour cells, turning them into necrotic cells.
Red box: Tumour cells proliferate; an \MOne-like macrophage adopts a more \MTwo-like phenotype after being exposed to TGF-$\beta$ near the tumour; an \MTwo-like macrophage moves towards blood vessels, attracting two tumour cells with it.
}
\label{fig:simulation}
\end{figure}

Macrophage phenotype is modelled as a continuous subcellular variable $p_i \in [0, 1]$:
$p_i = 0$ represents an anti-tumour `\MOne' phenotype, and $p_i = 1$ represents a pro-tumour `\MTwo' phenotype, whilst a macrophage with an intermediate phenotype exhibits intermediate behaviour.
As tumour cells release the diffusible chemoattractant \textit{colony stimulating factor-1} (CSF-1), \MOne-like macrophages with phenotype $p_i = 0$ extravasate from blood vessels and move towards tumour cells to kill them.
Under sufficient exposure to \textit{transforming growth factor beta} (TGF-$\beta$) in the tumour microenvironment, the macrophages adopt 
an \MTwo-like phenotype and migrate towards nearby blood vessels, guiding the migration of their neighbouring tumour cells. If sufficient numbers of tumour 
cells migrate with the $\MTwo$ macrophages all the way to a blood vessel, a {\em perivascular niche} of tumour cells may be formed. 
%

Once tumour cells accumulate and proliferate around blood vessels to form a perivascular niche, the likelihood of tumour intravasation and subsequent metastasis naturally increases. 
Indeed, such microenvironments are regarded as ``propulsion centres of the tumour'' \cite{niche}. There are various proposed methods for quantifying and studying shape of such cellular behaviours. For example, tumour cell counts and the minimum tumour-vessel distance provide a reasonable indicator of the presence of perivascular niches at the end of the simulation (see Fig~\ref{fig:summary}),
but subtle shape changes may allow for prediction at earlier stages before the tumour approaches the vessel, fragments, or gets eliminated. Other spatial statistics such as the pair-correlation function have been applied to study cellular heterogeneity and organisation \cite{bull_spatial_stats,Josh:ABM}. 

In recent years, there has been a wide range of research applying \textit{topological data analysis} (TDA) to biomedical data to quantify differences in shape and behaviour, as described in surveys such as \cite{amezquita2020, rabadan_blumberg}, and even, as we do in this work, to study an agent-based model of immune cell infiltration into a tumour-spheroid \cite{Vipond:MPH-landscape}. TDA is a growing field of computational mathematics that studies the shape of data. Standard \textit{persistent homology} (PH), the flagship technique, finds rich theoretical underpinnings in algebraic topology and statistics \cite{1PH:stability}. A discrete structure is built from the geometry of the data, and PH measures the evolution of topological features such as connected components, loops, and cavities as a single descriptor varies, often a measure of scale \cite{carlsson2009topology}\cite{ghrist2008barcodes}. As with any data analysis method, the modeller must make certain choices. Here, a series of choices propagate through the persistent homology pipeline: data $\rightarrow$ filtered complex $\rightarrow$ persistent homology $\rightarrow$ statistics and interpretation (e.g. see Figs.~\ref{fig:rips_pipeline},\ref{fig:radial_pipeline},\ref{fig:vineyard},\ref{fig:zigzag}). In practice, constructing the appropriate filtration is particularly important since different filtrations capture different geometric features of the data for direct analysis and vectorisation, which maps the persistence diagram into different vector spaces for classification and analysis \cite{stolz-pretzer2019a}.

Even in the simplest case of standard monotonic, one-dimensional PH, computation of persistent homology requires a choice of meaningful distance, function, or parameter on which to filter the data, which may or may not be obvious in context. For spatial data, natural choices include distances between data points (see Vietoris-Rips, Section \ref{sec:VR}) and distances of data from a fixed centre (see radial, Section \ref{sec:radial}), although there are many others, such as density, distance to measure, or any function built from an important feature of the data. The complexity of the natural topological space of persistence diagrams, on the other end of the pipeline, lacks inner products and unique averages, and requires a choice of vector representation to apply most statistics and classification methods \cite{ali2022survey}. We use the persistence image \cite{persistence-images} for all of our methods, as it tends to work well on real data, as the input for a logistic regression classifier.

As the data have multiple cell types, with both spatial and temporal features, we make several different choices of emphasis and compare the results. The widely used \textit{(Vietoris-) Rips filtration} (VR), which focuses on multi-scale data analysis, has been applied to snapshot dynamical systems data, including agent-based models of biological aggregations \cite{Topaz2015}. These methods compute VR persistence at each snapshot and track the variation of real-valued summary statistics as time evolves to create topological time series data. In the past few years, studies such as \cite{prostate, epithelial, neuroblastoma, lunghistology} have successfully applied VR features for machine-learning classification of real cell configurations, including histology slides of tumours in situ. Furthermore, VR features computed from multiple cell types have been used to classify spatial patterns arising from simulations of an agent-based model \cite{Dhananjay_heterogeneous}. Besides VR, more bespoke filtrations have been proposed in other applications. For example, the \textit{radial filtration} we use has been previously applied to neuronal trees and tumour vascular networks \cite{kanari,stolz2022}. In this filtration, distance from a central basepoint replaces pairwise distances of cell locations; this encodes the boundary tortuosity, components, and holes according to their radius and reach \cite{bubenik2020persistent}.


The {\em persistence vineyard} is another variant; it 
tracks the temporal variation of a VR complex directly in the persistence diagrams  \cite{vineyard}. Persistence vineyards have been recently applied to find regional anomalies of COVID-19 cases and functional connectivity of the brain \cite{covid-vineyard,fMRI-vineyard}, among other applications. They allow for an approximation of the natural continuity between topological disruption events.

Computing topological persistence over time using data continuity directly requires a different approach due to non-monotonicity of the corresponding complex. \textit{Zigzag persistence} \cite{Carlsson:zigzag} was developed theoretically alongside scale-parameter methods, and it tracks the evolution of individual features directly rather than the diagram as a whole. Zigzag PH has been applied to analyse complex systems \cite{corcoran2016} and to extract higher-dimensional structure from 2D images \cite{mata_zigzag}. However, computation restricted the application of zigzag persistence to smaller data sets until the recent algorithmic breakthrough in \citet{Carlsson:zigzag-computational}. Since then, zigzags have been applied to investigate dynamical systems \cite{tymochko2020} and time series data from coral reef models \cite{Rob:zigzag-coral}. 

 In this work, we implement two variations each of static and dynamic methods and apply them to spatial data from Bull and Byrne's agent-based model of tumour-immune interaction in order to compare their efficacy in predicting tumour elimination and escape. From the spatial locations of tumour cells, macrophages, and blood vessels, we demonstrate four approaches in total:
the Vietoris-Rips and radial filtrations quantifying the shape of the cell point cloud at individual timesteps, 
as well as the persistence vineyards and zigzag persistence tracking the evolution of cell clusters through the simulation.
We show that these techniques give complementary, interpretable spatio-temporal information, and we compare their predictive power by training a logistic regression classifier on each to detect patterns which predict the formation of perivascular niches in simulated data. We show significant improvements over benchmark statistics, including higher accuracy and earlier classification. Additionally, we interpret the logistic regression coefficients to identify qualitative shape features of pre-niche simulations.

\section{Bull and Byrne Model}

\begin{figure}[htbp]
\centering
\includegraphics[width=\textwidth]{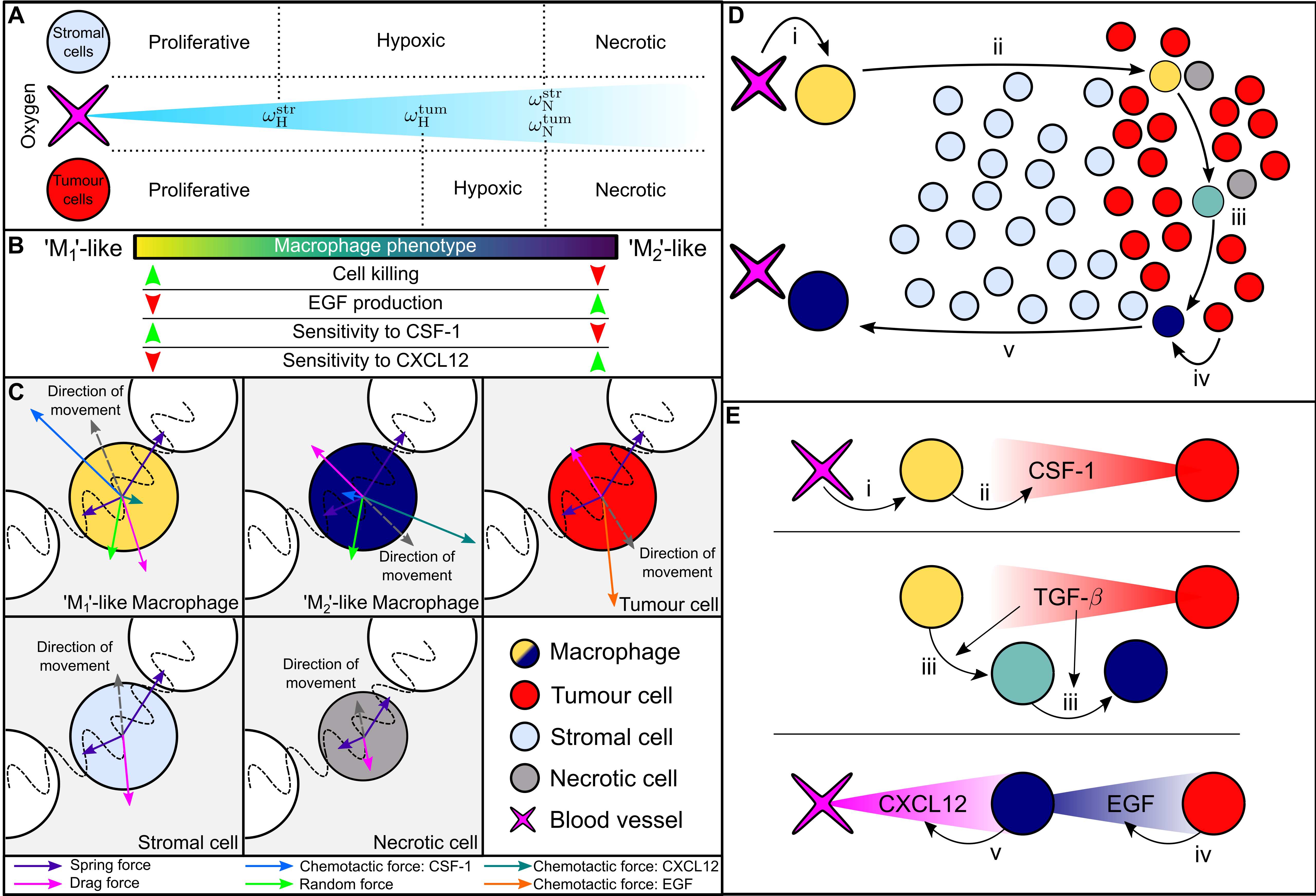}
\caption{{\bf Schematic of the ABM.}
Cell states and movement depend on the surrounding cells and local levels of multiple diffusible species. A: Stromal and tumour cells proliferate and die (via necrosis) at rates which depend on local oxygen levels (hypoxic thresholds). B: $\MOne$ macrophages 
are supplied to the tissue from blood vessels at a rate which depends on the CSF-1 concentration there. They migrate up spatial gradients of CSF-1 which is produced by the tumour cells and kill tumour cells on contact. As the $M_1$ macrophages get closer to the tumour, they are exposed to TGF-$\beta$ which drives them to a pro-tumour $M_2$ phenotype. $\MTwo$ macrophages are subject to chemotactic forces that direct them to move up spatial gradients in CXCL12, itself produced from blood vessel locations (by perivascular fibroblasts, not included in the model). The $\MTwo$ macrophages also produce EGF which acts as a chemoattractant for tumour cells, so that they move with the $\MTwo$ macrophages, towards neighbouring blood vessels.
C: Cell movement depends on the forces that act on a cell. These include chemotactic forces whose magnitude depend on a cell's sensitivity to spatial gradients in the diffusible species which, in turn, depend on cell type and phenotype.
D: Summary of the phases of macrophage-mediated tumour cell migration in our ABM.
i) \MOne{} macrophages extravasate from blood vessels in response to CSF-1. ii) \MOne{} macrophages migrate into the tumour mass in response to CSF-1, where they may kill tumour cells. iii) Exposure to TGF-$\beta$ causes macrophages to adopt an \MTwo{} phenotype. iv) \MTwo{} macrophages produce EGF, which acts as a chemoattractant for tumour cells. v) \MTwo{} macrophages migrate towards blood vessels, in response to CXCL12 gradients.
E: Schematic summarising the sources of CSF-1, TGF-$\beta$, EGF and CXCL12, and their interactions with cells, as described in steps i-v of panel D.
Reproduced from \cite{Josh:ABM}, Fig 2.
}
\label{fig:ABM_schematic}
\end{figure}

We briefly describe 
the two-dimensional, off-lattice agent-based model (ABM) of tumour-immune interactions developed by \citet{Josh:ABM} (see Figures~\ref{fig:simulation} and
~\ref{fig:ABM_schematic}). 
The model distinguishes macrophages, stromal, tumour and necrotic cells, and several diffusible species, including oxygen which is supplied by blood vessels located at fixed points. 

Each macrophage is associated with a subcellular variable or {\em phenotype} $p \in [0,1]$ which governs its behavior: anti-tumour macrophages, also called $\MOne$, have low $p$ values, {\em extravasate} from blood vessels, migrate toward tumour cells (up spatial gradients of the diffusible species CSF-1), and attempt to kill them on contact. Exposure to TGF-$\beta$, near the tumour, causes macrophages to transition to an $\MTwo$, or pro-tumour phenotype, which is associated with high values of $p$ ($p \approx 1$). Pro-tumour macrophages migrate towards blood vessels (up spatial gradients of the diffusible species CXCL12). Paracrine signalling between the $\MTwo$ macrophages and the tumour cells (mediated by the diffusible species EGF), causes the tumour cells to move with the $\MTwo$ macrophages, towards the blood vessels, where they intravasate and then metastasise to other parts of the body. 

 Stromal and tumour cells proliferate in response to oxygen supplied by the blood vessels. If they have sufficient oxygen, the cells proliferate, with tumour cells typically having a lower oxygen threshold for proliferation than stromal cells. If oxygen levels are too low for proliferation, or if a tumour cell is exposed to an $\MOne$ macrophage for too long, then it becomes necrotic. Stromal cells move primarily in response to physical forces exerted by surrounding cells, while tumour cells are also subject to chemotactic forces, which direct their movement up spatial gradients of EGF. Macrophages move in response to physical forces and explore their local environment via a random walk, as well as following chemotactic gradients of key diffusible species 
 (CSF-1 for $\MOne$ and CXCL12 for $\MTwo$ macrophages). 
 For further details of the ABM, and code to reproduce it within the Chaste (Cancer, Heart and Soft Tissue Environment) framework~\cite{chaste}, we refer the interested reader to \cite{Josh:ABM}.

\begin{figure}[h]
\includegraphics[width=\textwidth]{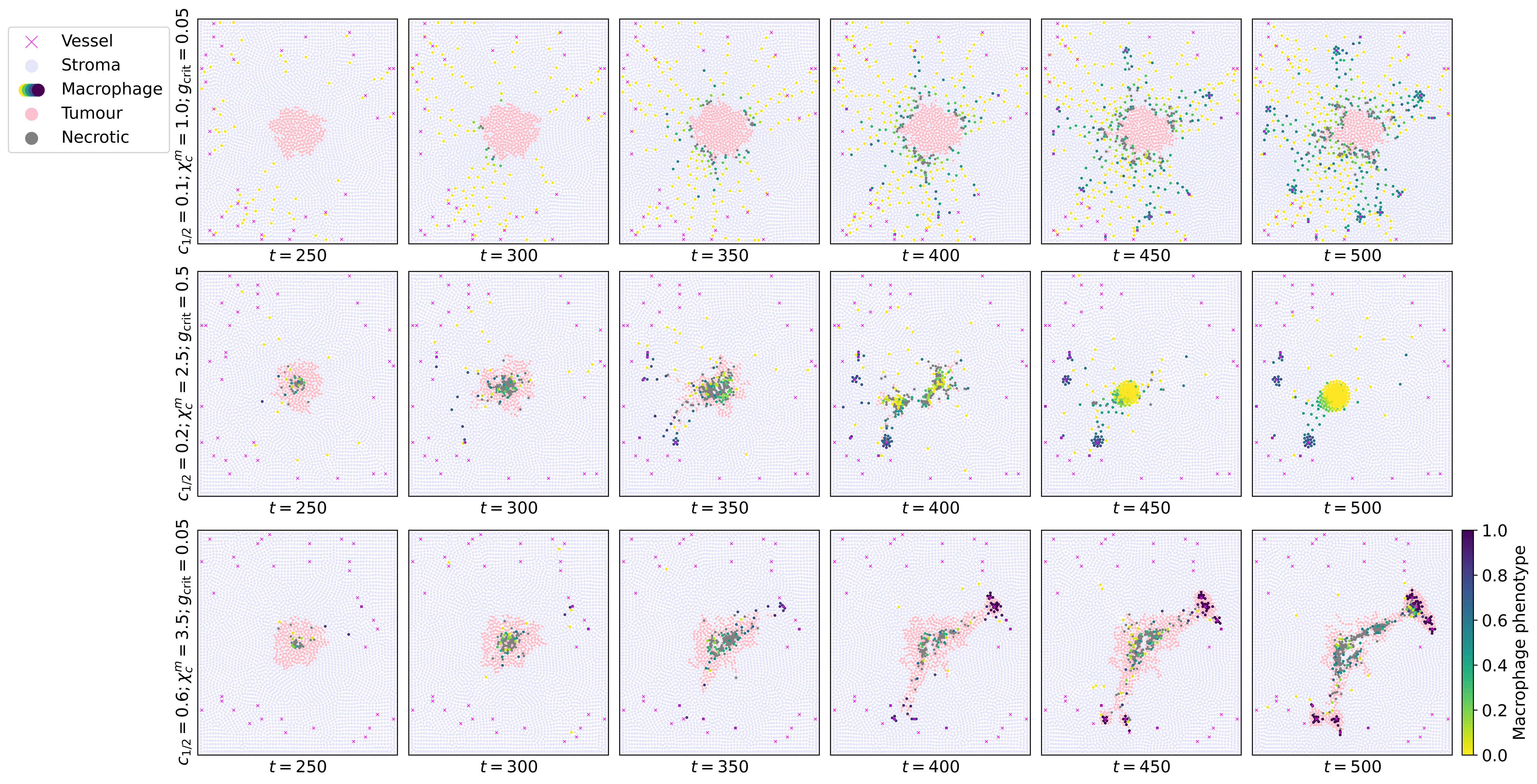}
\caption{{\bf Diverse patterns of tumour growth.}
Series of plots showing how the spatial patterns generated by the ABM change as we vary three parameters of interest: the extravasation rate $c_{1/2}$, the chemotactic sensitivity of $\MOne$ macrophages to CSF-1 $\chi_c^m$, and the critical TGF-$\beta$ threshold at which macrophages start to transition from an $\MOne$ to an $\MTwo$ phenotype $g_\mathrm{crit}$. The simulations illustrate three possible outcomes at $t = 500$.
Top:~Compact tumour mass in equilibrium; the dominant macrophage phenotype is \MOne.
Middle:~Tumour elimination; the dominant macrophage phenotype is \MOne.
Bottom:~Tumour escape; the dominant macrophage phenotype is \MTwo.
}
\label{fig:params}
\end{figure}

From the ABM, we generate $1485$ time series of tumour cells, macrophages and blood vessel locations, spanning $0$--$500$ hours and sampled every $10$ hours. 
We vary three parameters that regulate the behaviour of the macrophages:
 the extravasation rate $c_{1/2}$, the chemotactic sensitivity $\chi_c^m$ to spatial gradients in CSF-1, and the critical threshold $g_\mathrm{crit}$ of TGF-$\beta$ exposure levels above which a macrophage's phenotype $p_i$ increases at a constant rate. 
Fig~\ref{fig:params} illustrates the range of tumour growth patterns that the ABM exhibits. We observe that the {\MOne/\MTwo} ratio correlates with the qualitative outcome; this is also consistent with clinical evidence in colorectal and ovarian cancers
\cite{colorectal, ovarian}.

We characterise a \textit{perivascular niche} as a blood vessel with at least $10$ tumour cells within $5$ units of distance. Perivascular niches greatly increase the likelihood of metastasis, as the tumour cells can easily enter the blood vessels and be transported to other parts of the body.
With this definition, perivascular niches are present in $588$ (39.6\%) of the simulations at $t=500$. 

\section{Topological data analysis}
We now present the four TDA pipelines (Figs.~\ref{fig:rips_pipeline},\ref{fig:radial_pipeline}, \ref{fig:vineyard}, and \ref{fig:zigzag}), placing emphasis on intuition rather than rigour. For a comprehensive introduction to the theory of persistent homology, see \cite{carlsson2009topology} or \cite{ghrist2008barcodes};
we refer additionally to  \cite{persistence-images} for background on the persistence image vectorisation.  We first briefly describe the general TDA pipeline and then overview the four methods applied to this synthetic ABM data set.

We approximate the topology of data by first building a simplicial complex.
A \textit{simplicial complex} is a higher-dimensional generalisation of a network consisting of nodes and edges, but also triangles, tetrahedra, and higher-dimensional analogues of triangular solids. A {\em filtration} of a complex $K$ is a function $f:K\rightarrow \mathbb{R}$ such that each sub-level set $K_c := \{\sigma \in K : f(\sigma) \leq c\}$ is itself a simplicial complex. For each sequence of discrete values $c_1\leq c_2\leq \dots,\leq c_n$ we obtain a sequence of inclusion maps
$ K_{c_1}\subseteq K_{c_2}\subseteq \dots K_{c_n}.$

The simplicial homology $H_i(K)$ of a complex $K$ is an algebraic object encoding topological features of dimension $i$. The 0-dimensional homology $H_0$ is generated by connected components of the complex, and the 1-dimensional homology $H_1$ is generated by holes or loops. Persistent homology $\mathrm{PH}_i$ simultaneously computes the $i$-homology of an entire filtration, and encodes how $i$-dimensional features are created or destroyed over the course of the filtration:
in $K_{c_b}$, a new connected component or cluster might arise $(i=0)$, or a new hollow cycle of edges might form $(i=1)$;
in $K_{c_d}$, two connected components might merge into one, or a cycle might become triangulated and filled in.
The pair $(c_b, c_d)$ records the \textit{birth-death} values of the feature with respect to the filtration, and its \textit{persistence} is the lifespan $c_d - c_b$.
By the classic Structure Theorem, $\mathrm{PH}_i(K_\varepsilon)$ has a unique decomposition into intervals, which can be plotted as a barcode, or their endpoints plotted in $\mathbb{R}^2$ as a \textit{persistence diagram}.
In turn, the diagram can be vectorised into a \textit{persistence image} \cite{persistence-images} by placing a Gaussian kernel over all points.

\subsection{Vietoris-Rips filtration.}\label{sec:VR}
For a finite point cloud $X \stackrel{\text{fin}}{\subseteq} \mathbb{R}^2$, and radius parameter $\varepsilon \geq 0$, the \textit{Vietoris-Rips complex} $K_\varepsilon$ associated to $X$ is given by:
\begin{itemize}
\item a node ($0$-simplex) for each element $x\in X$;
\item an edge ($1$-simplex) $\{x, y\}$ when $x, y \in X$ satisfy $d(x, y) \leq \varepsilon$;
\item a triangle ($2$-simplex) $\{x, y, z\}$ if $x, y, z \in X$ satisfy $d(x, y), d(y, z), d(z, x) \leq \varepsilon$; 
\item more generally, an $n$-simplex $\{x_0, x_1, \dots, x_n\}$ when $d(x_i,x_j)\leq\varepsilon$ for all $x_i,x_j$.
\end{itemize}
We note that although our data lies in the real 2-dimensional plane, $K_\varepsilon$ is a discrete combinatorial object which can contain components of any dimension up to $|X|$. 

\begin{figure}[h]
\includegraphics[width=\textwidth]{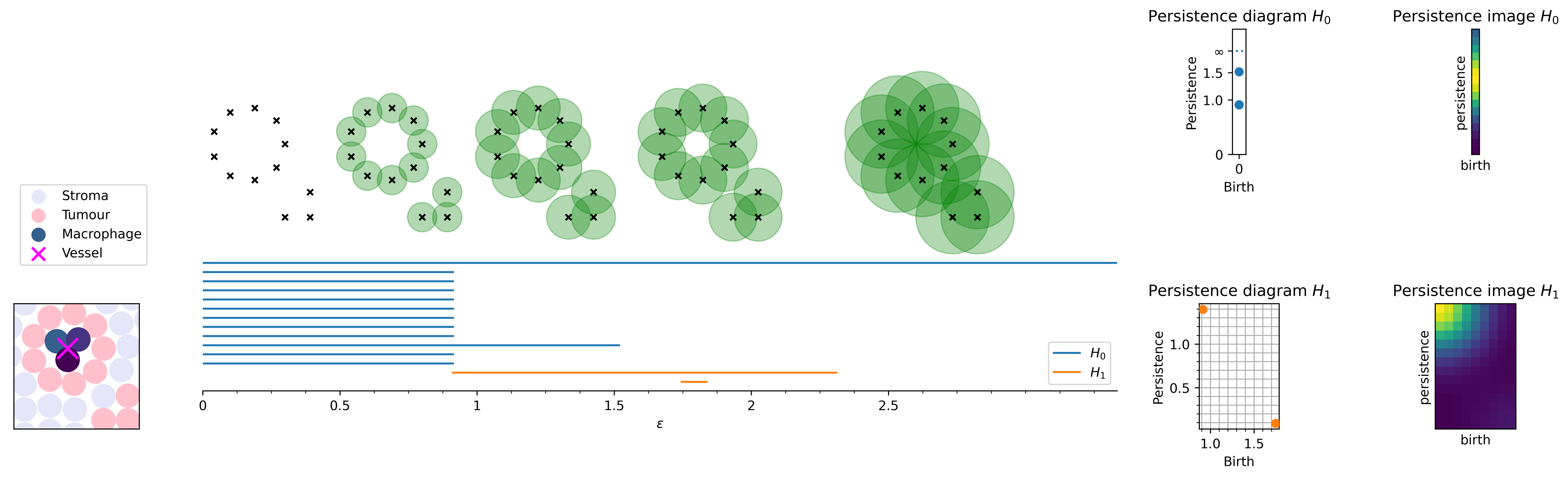}
\caption{{\bf Pipeline for the Vietoris-Rips PH computation.}
Left: Spatial cell data. 
Centre top: Neighbourhoods of increasing radius around tumour coordinates $T$. 
Centre bottom: The barcode $\mathrm{PH}_0(K_\varepsilon)$, blue, has a bar for each component of $K_{\varepsilon_t}$ spanning the time interval on which it can be seen in the filtration $K_\varepsilon$; $\mathrm{PH}_1(K_\varepsilon)$, orange, contains a bar for each hole in $K_\varepsilon$ over its corresponding interval. 
Right: The left endpoint and length (called the ``birth" and ``persistence", resp.) of each bar are plotted in $\mathbb{R}^2$ to form the persistence diagram. A Gaussian kernel, weighted by the persistence value, is placed over every point in the diagram to form the corresponding persistence images $\RipsIm_0(T)$ and $\RipsIm_1(T)$. }
\label{fig:rips_pipeline}
\end{figure}


 Given a finite sequence of radii $\varepsilon_0 \leq \varepsilon_1 \leq \dots \leq \varepsilon_n$, we obtain a nested sequence of simplicial complexes 
$
K_{\varepsilon_0} \subseteq K_{\varepsilon_1} \subseteq \dots \subseteq K_{\varepsilon_n},
$
since simplices are added monotonically as $\varepsilon$ grows. This is the \textit{Vietoris-Rips} filtration. The birth and death of each homological feature indicates its scale: at small $\varepsilon$, only neighbouring points are connected by simplices, while at large $\varepsilon$, more global features appear.
 
 This pipeline is summarised in Figure \ref{fig:rips_pipeline}, and we will denote the output image by $\RipsIm_i(X)$, where $i$ is the dimension and $X$ is a static point cloud of macrophages $M$ or tumour cells $T$ at a specific time step. We use \texttt{Ripser} \cite{Bauer2021Ripser} to compute $\mathrm{PH}_i(K_\varepsilon)$ for $i=0,1$, and \texttt{Persim} to generate the images $\RipsIm_0$ and $\RipsIm_1$ from these outputs; we use the default linear weighting by persistence for Gaussian intensity.

\subsection{Radial filtration.}\label{sec:radial}
\begin{figure}[h]
\centering
\includegraphics[width=0.65\textwidth]{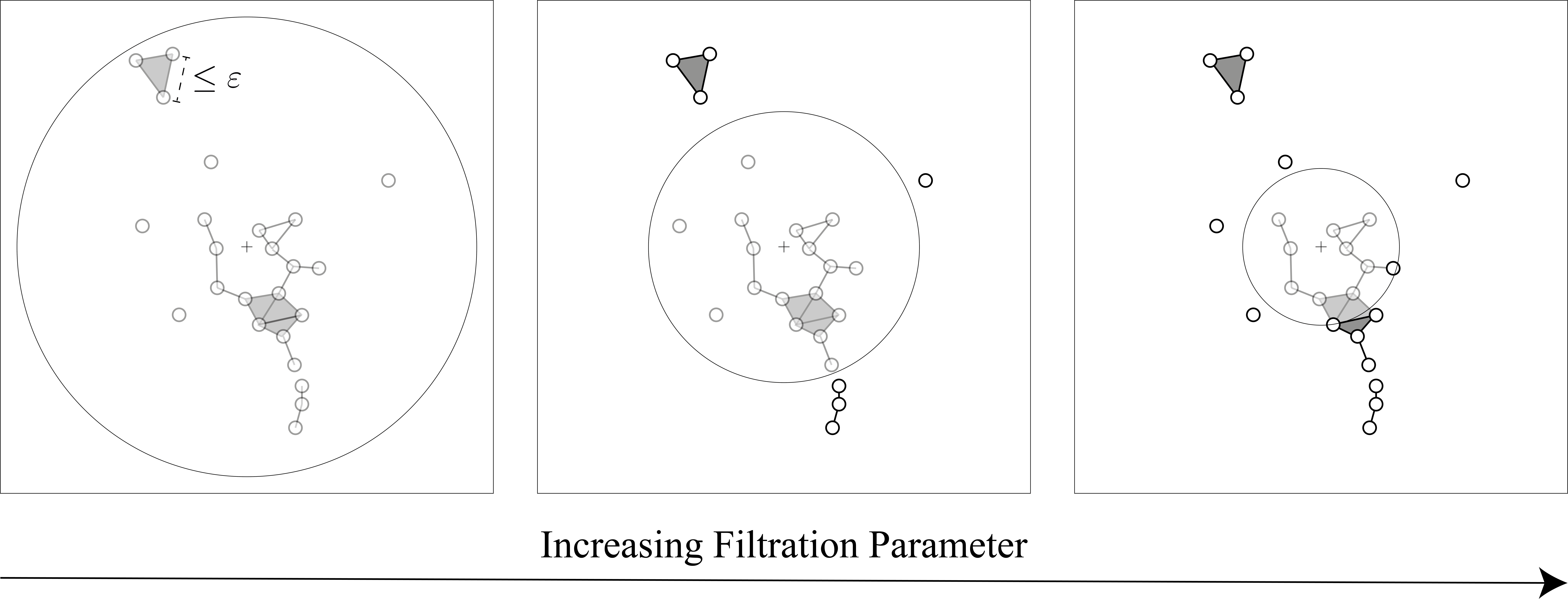}
\includegraphics[width=\textwidth]{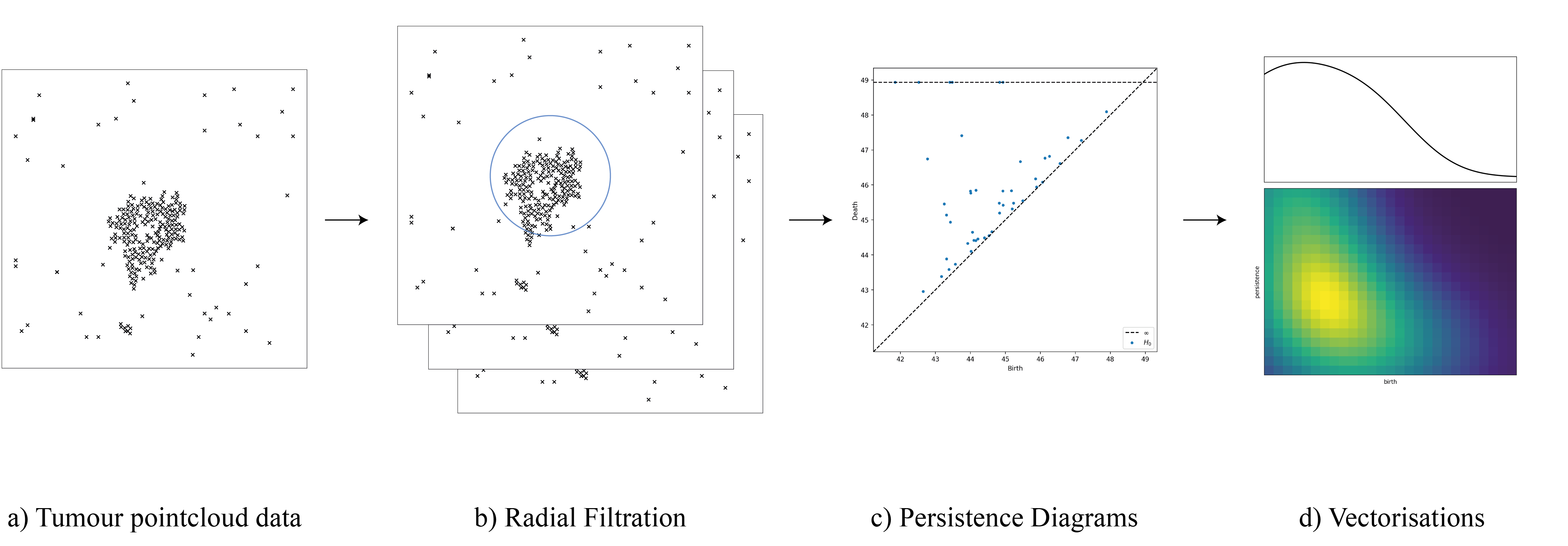}
\caption{{\bf Pipeline for the radial PH computation for tumour cells.} 
a: Input is a pointcloud of cells, and basepoint $\mu$ in the centre of the simulation. b, shown in more detail at top: A simplicial complex is built by adding a point for each tumour cell, connecting neighbouring cells whenever two tumour cells have distance at most $\varepsilon$, and adding higher simplices whenever their outline is included. This complex is then filtered by radius, including the furthest points at low filtration parameters and more central points as the parameter increases. c: The persistent homology of the filtration is computed. d: The resulting diagram is converted to a two-part persistence image; a 2D image representing finite features and a 1D function representing the connected components of the full complex and their distance from the $\mu$. 
}
\label{fig:radial_pipeline}
\end{figure}

For a finite point cloud $X \stackrel{\text{fin}}{\subseteq} \mathbb{R}^2$, we fix a linkage distance $\varepsilon$ and construct a (Vietoris-Rips) simplicial complex $K_{\varepsilon}$ on the point locations. This is used as the underlying complex for the filtration. 
As shown in Figure \ref{fig:radial_pipeline}, at parameter $w \geq 0$, the simplicial complex $K_\varepsilon^{(w)}$ is defined
$$K_{\varepsilon}^{(w)}: = \left\lbrace\sigma \in K_\varepsilon \hspace{0.5em}:\hspace{0.5em} 35-||x-\mu|| \leq w \hspace{1em}\forall x\in \sigma \right\rbrace,$$
where $\mu$ is a fixed basepoint in the middle of the point cloud, $||\cdot||$ is the standard Euclidean norm, and $||\mu-x||< 35$ for all points $x\in X$. 
In this way, $K_{\varepsilon}^{(w)}$ includes the points which are over $35-w$ away from the base point as 0-simplices, as well as all edges, triangles, and higher simplices between these points. Given $0 = w_0 \leq w_1 \leq \dots \leq w_n = 35$, we obtain a filtration
\[
\emptyset = K_{\varepsilon}^{(w_0)} \subseteq K_{\varepsilon}^{(w_1)} \subseteq \dots \subseteq K_{\varepsilon}^{(w_n)} = K_\varepsilon
\]
which we call the {\em radial filtration}.

We use \texttt{GUDHI} to construct this filtration and to compute its PH, and \texttt{Persim} to compute persistence images, as before. However, standard persistence images exclude features of {\em infinite} persistence: these are homological features that are still present at the final filtration step $K_\varepsilon$, and therefore are said to ``never die". We reintroduce these features, which are parametrised in one dimension by their birth radius.

The infinite features in $H_0$ correspond to the connected components of the underlying simplicial complex $K_{\varepsilon}$. The infinite features are vectorised as a real-valued function over the range of radii by taking a sum of 1D Gaussian kernels centred at their birth times. This 1D persistence image is appended as a vector to the persistence image obtained from the finite points. The resulting image is called $\mathrm{PI}^\mathrm{rad}$.

\subsection{Time-dependent topological analyses}
We now turn to the first of two time-dependent analyses.
Given a finite point cloud $X_t \stackrel{\text{fin}}{\subseteq} \mathbb{R}^2$, that is continuously changing location over time, we track how and when clusters form, persist and die. Rather than study a single time snapshot, we use the cumulative trajectory from many timesteps.


\subsubsection{Persistence vineyard.}\label{sec:vineyard}

A seemingly naïve method of tracking evolution over time is to compute $\PH_i(\operatorname{VR}(X_t))$, the $\PH$ of the Vietoris-Rips filtration, at each individual time step $t$. 
The Vietoris-Rips stability theorem \cite{persistence-diagram:stability, stability_Chazal} states that a small perturbation of the input point cloud (in Gromov-Hausdorff distance) only elicits a small change in the resultant persistence diagram. If the time series is continuous with respect to this distance, then,
it follows that points in the persistence diagram far from the diagonal vary continuously in $t$. 

\begin{figure}[h]
    \centering
    \includegraphics[width=0.65\textwidth]{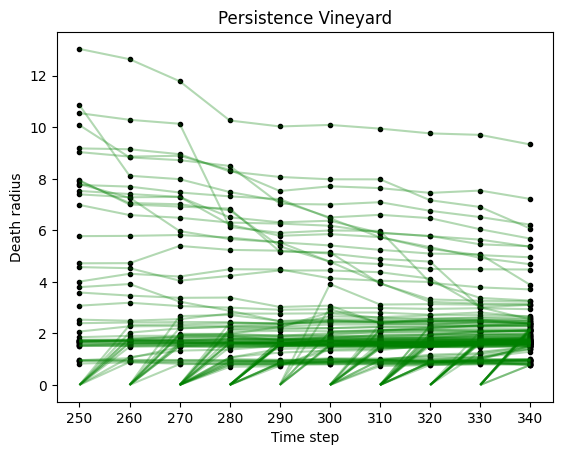}
    \caption{An example of a persistence vineyard approximation in dimension 0, with each vine (in green) tracking the change in death radius over time of a single feature. Each feature is generated by a single cell. At radius 0, each cell is its own feature. As the radius grows, a feature develops as other cells cluster around it, and die as the group joins a larger cluster. Here the diagrams are matched directly to illustrate continuity - this step is omitted in our methods.}
    \label{fig:vineyard}
\end{figure}

When the persistence diagrams are stacked serially, with time as a third axis, in addition to birth and death, the points in the persistence diagrams trace out smooth curves (or when discrete time steps are taken, a piecewise-linear approximation) called \textit{vines} (see Fig.~ \ref{fig:vineyard}). 

Persistence vineyards do not admit an obvious vectorisation, but we can encode vines implicitly in the persistence image by plotting the aligned Rips diagrams and allowing the Gaussian kernel to spread across layers. As any point $(\varepsilon_b, \varepsilon_d)$ in $\operatorname{PH}_0$ of the filtration $K_{\varepsilon}^{(t)}$ at time step $t$ is born at $\varepsilon_b = 0$, the \textit{vineyard} of vines lies entirely in the two-dimensional plane $\varepsilon_b=0$.
Therefore, we do not explicitly extract the vines;
rather, by abusing notation, we can consider $(t, \varepsilon_d)$ as a birth-persistence pair and thus obtain a persistence image.
We discard the single point $(\varepsilon_b, \varepsilon_d) = (0, \infty)$ at all time steps since it does not contribute additional information in the Rips filtration.

Representing the corresponding stacked persistence diagram in dimension 0 with a persistence heat-kernel image as in previous sections, we denote the result $\vinIm$.

\subsubsection{Zigzag persistence.}\label{sec:zigzag}

\begin{figure}[h]
    \centering
    \includegraphics[width=0.9\textwidth]{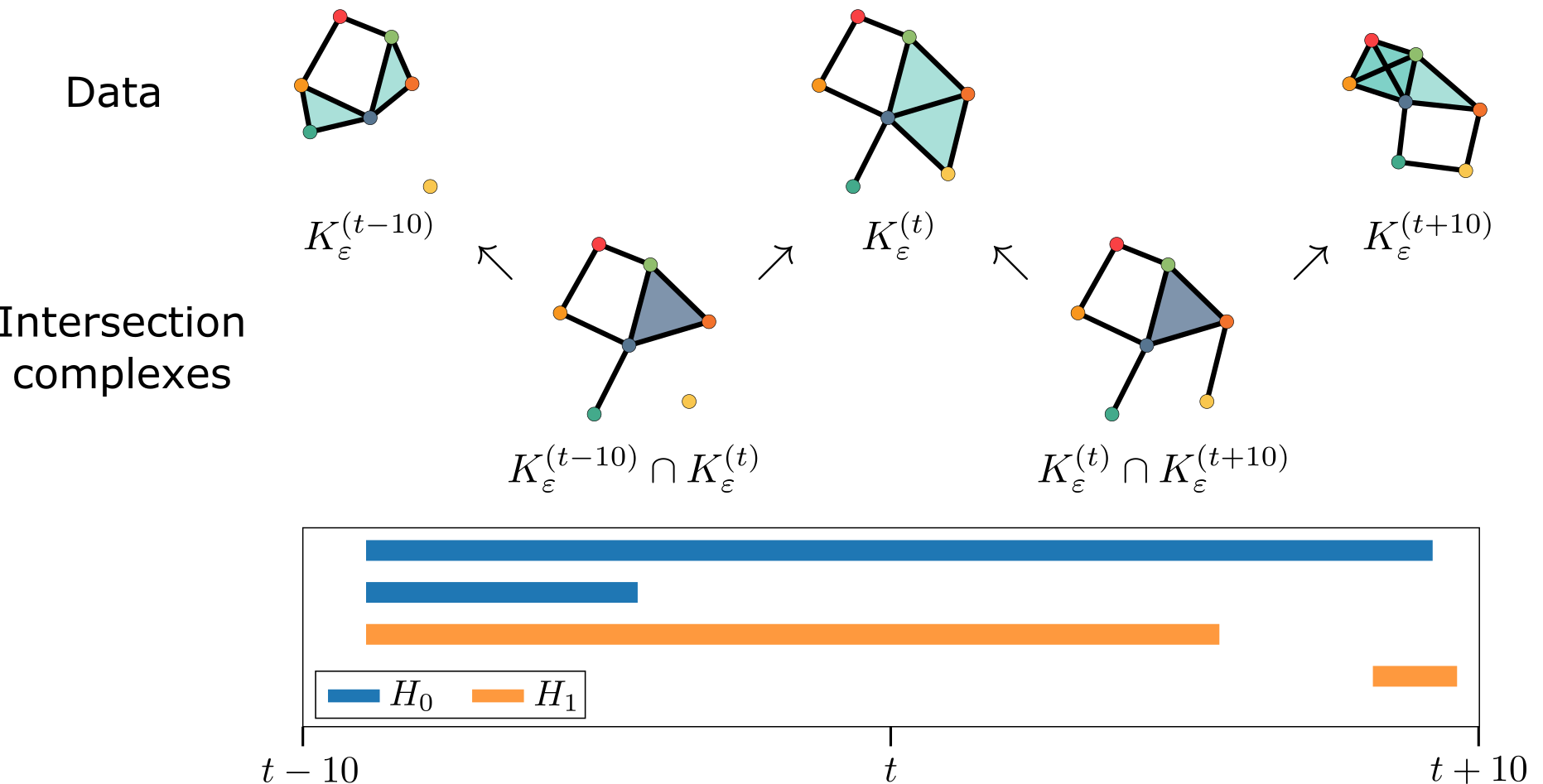}
    \caption{\textbf{Zigzag PH detects persistent cycles across a sequence of simplicial complexes and their pairwise intersections.} Top: Given a sequence of simplicial complexes, say $K^{(t-10)}_{\varepsilon}$, $K^{(t)}_{\varepsilon}$, and $K^{(t+10)}_{\varepsilon}$, zigzag PH first constructs intersected complexes for each consecutive pair, resulting in a zigzag filtration. Bottom: The barcode associated with the zigzag filtration. }
    \label{fig:zigzag}
\end{figure}

One main drawback of the persistence vineyard is that it typically identifies features across consecutive time steps based on an imposed matching of topological features, where we assume that features $(b,d)$ and $(b',d')$ belonging to consecutive time steps represent the evolution of the same feature if they have a similar birth and death time relative to other features. If the time steps are sufficiently dense, this may be accurate but, in general, this assumption does not hold. Ideally, we would like to incorporate knowledge of which cells are co-located in a topological feature to determine if a feature persists from one time step to the next, or if it dies between time steps while a similar-looking feature is born. We will accomplish this via a {\em zigzag }filtration (Figure~\ref{fig:zigzag}). 

In order to track the actual, rather than estimated, persistence over time of homological features, we implement a non-monotonic filtration called a {\em zigzag} \cite{Carlsson:zigzag}. Zigzag filtrations exploit simplicial maps to compute the persistent homology of a sequence of simplicial complexes which is not strictly growing, but can both add and remove simplices. This allows us to compute persistence over time and capture the topological dynamics. However, it requires some pre-processing: we must alter our sequence so that each map is either an inclusion or a restriction. To do this, we construct the {\em intersection complex} $K_\varepsilon^t\cap K^{t+10}_\varepsilon$, and compute the persistent homology of the sequence as shown in Fig.~\ref{fig:zigzag}, using the natural inclusion maps of the intersection. This produces the \textit{zigzag filtration}
\[ \cdots \rightarrow K^{(t - 10)}_{\varepsilon} \leftarrow K^{(t - 10)}_{\varepsilon} \cap K^{(t)}_{\varepsilon}  \rightarrow K^{(t)}_{\varepsilon} \leftarrow K^{(t)}_{\varepsilon} \cap K^{(t + 10)}_{\varepsilon} \rightarrow K^{(t+10)}_{\varepsilon} \leftarrow \cdots. \]
The corresponding zigzag persistence module 
\[ \cdots \rightarrow PH_i(K^{(t - 10)}_{\varepsilon}) \leftarrow PH_i(K^{(t - 10)}_{\varepsilon} \cap K^{(t)}_{\varepsilon}  \rightarrow K^{(t)}_{\varepsilon}) \leftarrow PH_i(K^{(t)}_{\varepsilon} \cap K^{(t + 10)}_{\varepsilon}) \rightarrow PH_i(K^{(t+10)}_{\varepsilon}) \leftarrow \cdots \]
enjoys the same unique decomposition \cite{zigzag-decomposition} and stability \cite{zigzag-stability} properties as ordinary PH for all $i$. As a result, we can compute a \textit{zigzag barcode} for the zigzag persistence module (see Fig.~\ref{fig:zigzag}). We vectorise the persistent homology of the zigzag filtration with a persistence image, in exactly the same manner as Section \ref{sec:VR}. We call the result $\zzIm$.

\subsubsection{Comparing vineyards and zigzags}

Lastly, we highlight the different information provided by vineyards and zigzags. As previously discussed, in the vineyard, points in the persistence diagram represent multi-scale features connected by topological similarity over time. In the zigzag, a scale of neighbour connection must be fixed, similar to the radial filtration, and each simplex is tracked over time. This leads to slightly different dynamics, as the vineyard may detect spatial continuity that the zigzag does not, and the zigzag may detect temporal continuity that the vineyard does not.

\subsection{Classification}\label{sec:classification}
We use a $10$-time-repeated, stratified $5$-fold cross-validated logistic regression classifier to evaluate membership into two classes of tumour: those that will go on to form perivascular niches by time 500, and those that do not.
We benchmark their performance against the cell counts, the tumour-vessel distance, and the {\MOne/\MTwo} ratio, since state-of-the-art imaging techniques can provide information about the phenotype and the spatial position of individual cells.
Logistic regression also enables us to perform inverse analysis; using the regression coefficients, we can identify those regions of the persistence image which are most indicative of class membership. For example, high contrast at low persistence values in $\RipsIm_0$ indicates that the number of cell clusters at low single-linkage radii differs significantly between the groups, and the high contrast along the diagonal of the zigzag filtration regression coefficients indicates that features born at any time and lasting through to the final time step are most significant.  

\section{Results and discussion}


In this work, we apply four TDA methods to analyse patterns of tumour and macrophage cell locations: (1) Vietoris-Rips persistence images $\RipsIm_0(T),\RipsIm_1(T),\RipsIm_0(M),\RipsIm_1(M)$ of tumour\\ and macrophage cell locations in dimensions 0 and 1; 
(2) radial persistence images $\radIm$ of tumour cells in dim.\ 0;
    (3) zigzag persistence images $\zzIm$ of macrophages dim.\ 0;  and 
    (4) persistence vineyard images $\vinIm$ for macrophages in dim.\ 0. 
The first two,  
$\RipsIm_i$ and $\radIm$, are static data analysis methods at time $t$, while $\zzIm$ and $\vinIm$ analyse dynamics over a time period preceding $t$. The dimension 0 images track connected components over different positions, scales, or time. The dimension 1 images record the scale of holes in the cell clusters.

We describe the application and interpretation of the four topological pipelines on the agent-based model simulation dataset which contains 1485 time series data of tumour cells, macrophages, and blood vessel locations, with varying extravasation rate, chemotactic sensitivity, and critical TGF-$\beta$ threshold (see Fig.~\ref{fig:params}), spanning 0-500 hours. We analyse samples at time steps of 50 hours, beginning at 250 hours. For static methods, the current time step is analysed, while for dynamic methods, time steps of 10 hours for the preceding 100 hours are combined.

We assign a class label of 1 to time series that exhibit at least one perivascular niche at time 500 hours, and assign 0 otherwise.
For each persistence image, we train a logistic regression classifier as described in Section~\ref{sec:classification} and determine its ability to predict whether a given simulation time series will go on to form a perivascular niche. In Sections \ref{sec:VR_results}, \ref{sec:rad_results}, and \ref{sec:time_results}, corresponding to Figures \ref{fig:images_and_lr_rips}, \ref{fig:radial-results}, and \ref{fig:images_and_lr_time}, respectively, we examine the logistic regression coefficients and overall class differences between the images. This allows us to identify and interpret the regions of the persistence images that are most strongly linked (positively or negatively) to the future development of perivascular niches, and to determine what topological features they represent.
In Section \ref{sec:accuracies}, we present the accuracy scores given by the classifier at different time steps for each method and compare the performance of topological summaries with classic heuristics and with each other. 


\subsection{Vietoris-Rips persistence identifies tumour clusters and immune infiltration}\label{sec:VR_results}

\begin{figure}[ht]
\includegraphics[width=0.9\textwidth]{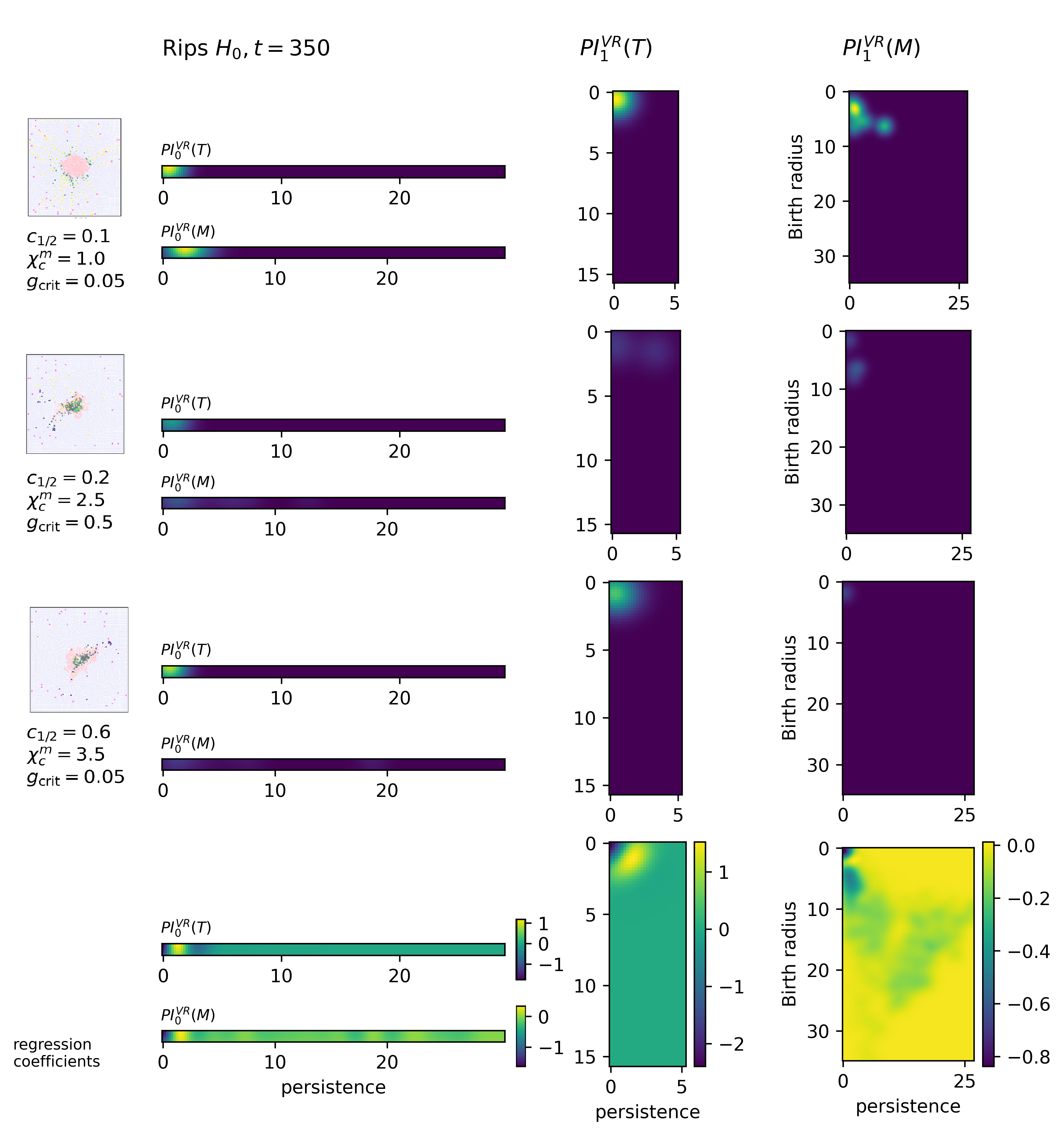}
\caption{{\bf Persistent images of VR filtrations.}
The top three rows correspond to the parameter values shown in Fig.~\ref{fig:params} at time $t = 350$. Each of the top three rows depict persistence images $\RipsIm_0(T),\RipsIm_1(T),\RipsIm_0(M),\RipsIm_1(M)$. 1st row: A future equilibrium case. 2nd row: Elimination case. 3rd row: Escape case. Last row: Regression coefficients identify regions of the persistence image that are most important for distinguishing the classes.}
\label{fig:images_and_lr_rips}
\end{figure}

Vietoris-Rips persistence is used to analyse the pairwise distances between point clouds of tumour cells $T$ and point clouds of macrophages $M$. We compute persistent homology of the Vietoris-Rips filtration in dimensions 0 and 1, take the persistence images of the homological features, and denote the result $\RipsIm_i(X),$ where $i\in \{0,1\}$ is the homological dimension and $X \in \{M,T\}$ is the cell type. 
$\RipsIm_0$ provides a multi-scale measure of the number, size, and proximity of the clusters of a particular cell type, whereas $\RipsIm_1$ measures the number, scale, and enclosure size of cavities inside those cell clusters. The logistic regression coefficients and sample persistence images from the Vietoris-Rips complexes are shown in Fig.~ \ref{fig:images_and_lr_rips}. 

\subsubsection{Vietoris-Rips of tumour cell analysis}

\paragraph{Tumour components --\boldmath$\RipsIm_0(T)$.}
For tumour cells, $\RipsIm_0(T)$ tracks the clusters of tumour cells and their proximity. Unless the tumour is separated into multiple distinct components, $\RipsIm_0(T)$ will have persistence only slightly higher than the cell radius $0.5$. A feature of higher persistence $p$ indicates a separate component at distance $p$ from the rest of the tumour. 

The logistic regression coefficients give some insight into the differences between the niche and control cases.  From Figure~\ref{fig:images_and_lr_rips}, we can see that the hallmark of the niche classes is components of persistence $1-2$, with non-niche cases showing more features of both smaller (0-1) and larger (2-4) persistence. 
The smaller features are likely individual tumour cells, which die around the optimal linking radius of $0.7$ found in the $\epsilon$-sweep for the radial filtration (See Fig. \ref{fig:eps_rad}). The features of higher persistence are disconnected from the body of the tumour, with at least one non-tumour cell in between. This can occur during elimination, or in the process of migration toward a blood vessel, although the results indicate that disconnection is more characteristic of non-niche cases.

The logistic regression classifier trained on $\RipsIm_0(T)$ outperforms the classifier trained on tumour cell count, which shows that the number and relative distance of tumour components differs meaningfully between the classes. However, $\RipsIm_0(T)$ does not predict perivascular niches as well as other topological methods (see Fig. \ref{fig:summary}). While fragmentation of the tumours occurs in elimination cases and  equilibrium cases, it may also occur during the process of escape.

\paragraph{Tumour holes --\boldmath${\RipsIm_1(T)}$.}
$\RipsIm_1(T)$, which represents cavities inside tumours, shows  improved performance of the classifier. The logistic regression coefficients (see Fig. \ref{fig:images_and_lr_rips}) contain a dark spot at $(0,0)$ which is associated with the non-niche class. These are cycles of tumour cells with very small spacing in between and a small loop radius so that the interior is only one or two cells wide. The bright yellow spot near $(1.5,1.5)$, which is correlated with the niche cases, represent loops of tumour cells that have a linking radius of 0-2 cells in between and an interior radius around two cells wide, or four cells in diameter. This pattern can be seen clearly in the examples shown, with a larger and more pronounced tumour hole in the escape case, a looser tumour loop in the elimination case, and a high number of loops of very small persistence in the equilibrium case.

While the coefficients do not tell us which other types of cells are in the interior of the cavity, an inspection of the simulations shows tumour loops around macrophages and around a single blood vessel. Both scenarios are associated primarily and frequently with simulations that give rise to escape, which we hypothesise drives the performance of classifiers using $\RipsIm_1(T)$ as input. The results also suggest that the number of infiltrated macrophages may play an important role in differentiating between the elimination and escape case at early time steps.

\paragraph{Macrophage clusters --\boldmath$\RipsIm_0(M)$.}
Macrophages have very different homological profiles. For both training classes, macrophage clusters form at a higher radius than tumour cells, and high-persistence features have higher weight in classification. However, the logistic regression coefficients of $\RipsIm_0(M)$ show a similar pattern to $\RipsIm_0(T)$: a positive correlation of features with persistence 1-2 in the niche case, with all other persistence values linked with lower likelihood of niche formation (see Fig. \ref{fig:images_and_lr_rips}). 
%

The logistic regression coefficients indicate that closely spaced, but not packed, macrophages may be a signature of the niche case, while the presence of densely-packed or diffuse clouds of macrophages predict non-niche behaviour. The higher linkage radius of macrophage components was also seen in our $\epsilon$ sweeps for the radial and zigzag filtrations. We think this is likely because macrophages move independently and do not proliferate, so they are less likely than tumour cells to form densely-packed clusters. Classifiers trained on $\RipsIm_0(M)$ outperform those trained on macrophage phenotype ratio and macrophage count (see Fig. \ref{fig:summary}), at timesteps up to 400 hours. Based on our tests, $\RipsIm_0(M)$ predicts perivascular niche formation up to 150 hours earlier than macrophage count or phenotype ratio. As the number of features in $\RipsIm_0(M)$ coincides with the total number of macrophages, this means that the spatial clustering of macrophages encodes useful information above and beyond the macrophage count. 

%
%

%
\paragraph{Macrophage holes --\boldmath$\RipsIm_1(M)$}
Interestingly, as shown in Figure \ref{fig:images_and_lr_rips}, the presence of any feature in $\RipsIm_1(M)$ is correlated with decreased likelihood of niche formation, at approximately the same accuracy levels as phenotype ratio $\MOne/\MTwo$ (Fig. \ref{fig:summary}).

In the example images, we see some patterns of macrophage behaviour which may be driving this result. 
$\RipsIm_1(M)$ has a dominant feature in the equilibrium case, as the macrophages assemble around the perimeter of the tumour, with the tumour mass enforcing a large cavity in the macrophage point cloud. In the elimination case, tumour fragmentation can create multiple macrophage voids as each tumour fragment become surrounded by macrophages. Additionally, extravasation may cause macrophage loops surrounding blood vessels. In general, however, we expect the majority of macrophage loops to contain tumour cells in the interior.

Based on the examples and the coefficients, we may also ascribe some of these results to the differing behaviour of $\MOne$-like and $\MTwo$-like macrophages, which may in part explain why the  accuracy level is so similar to phenotype ratio. $\MOne$ macrophages will remain on the boundary of the tumour, accumulating over time to form more well-defined loops around tumour components, while $\MTwo$ macrophages will return to the blood vessel.







\subsection{Radial persistence characterises tumour tortuosity}\label{sec:rad_results}

The \textit{radial filtration}, as shown in Figure \ref{fig:radial_pipeline}, captures the tortuosity and fragmentation of the tumour. Tumour cells outside radius $r$ are clustered and tracked as $r$ decreases, including cells nearer and nearer to the centre, noting when clusters merge.
The persistent homology $H_0$ therefore measures the variability 
of the outer boundary (e.g. tortuosity) of the tumour and the overall compactness of the central tumour. The infinite features of $H_0$, vectorised separately and appended as per Section \ref{sec:radial}, represent connected components in the underlying tumour complex.

\begin{figure}[ht]
    \centering
    \includegraphics[width=0.6\textwidth]{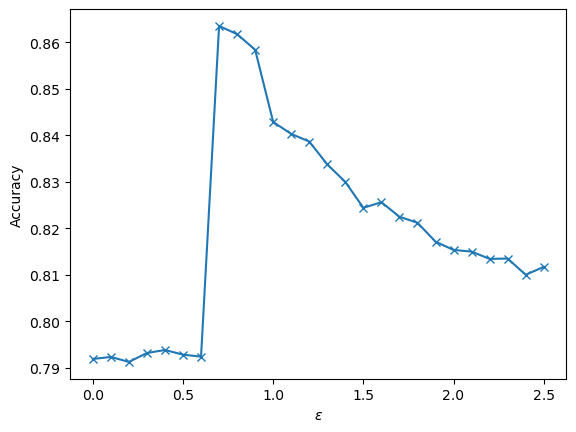}
    \caption{A coarse epsilon sweep to find the optimal connection radius of tumour component, where 1.0 is the diameter of each cell in the simulation. This gives some additional insight into the proximity of tumour cells and what constitutes a significant distance.}
    \label{fig:eps_rad}
\end{figure}

\paragraph{Linking radius}
We choose the cell neighbour distance $\varepsilon=0.7$ based on a coarse parameter sweep, as shown in Figure \ref{fig:eps_rad}. This linking radius was seen to maximise the classification accuracy of the filtration.

The optimal linkage distance $\varepsilon = 0.7$ is only slightly greater than the cell radius of $0.5$, indicating that tumour cells in the same component tend to be quite close together, and that tumour cells spaced more than $0.4$, boundary to boundary, from the main tumour may be considered separated for the purposes of classification. 
In our model, the central basepoint is aligned for all simulations at \((25,25)\), where the tumour is initially located. In more general cases, including with real data, a fixed basepoint cannot be assumed. We propose the centre of the tumour mass as a basepoint in these cases. See Section \ref{sec:real-data} for more discussion.

\begin{figure}[ht]
    \centering
    \includegraphics[scale=0.6]{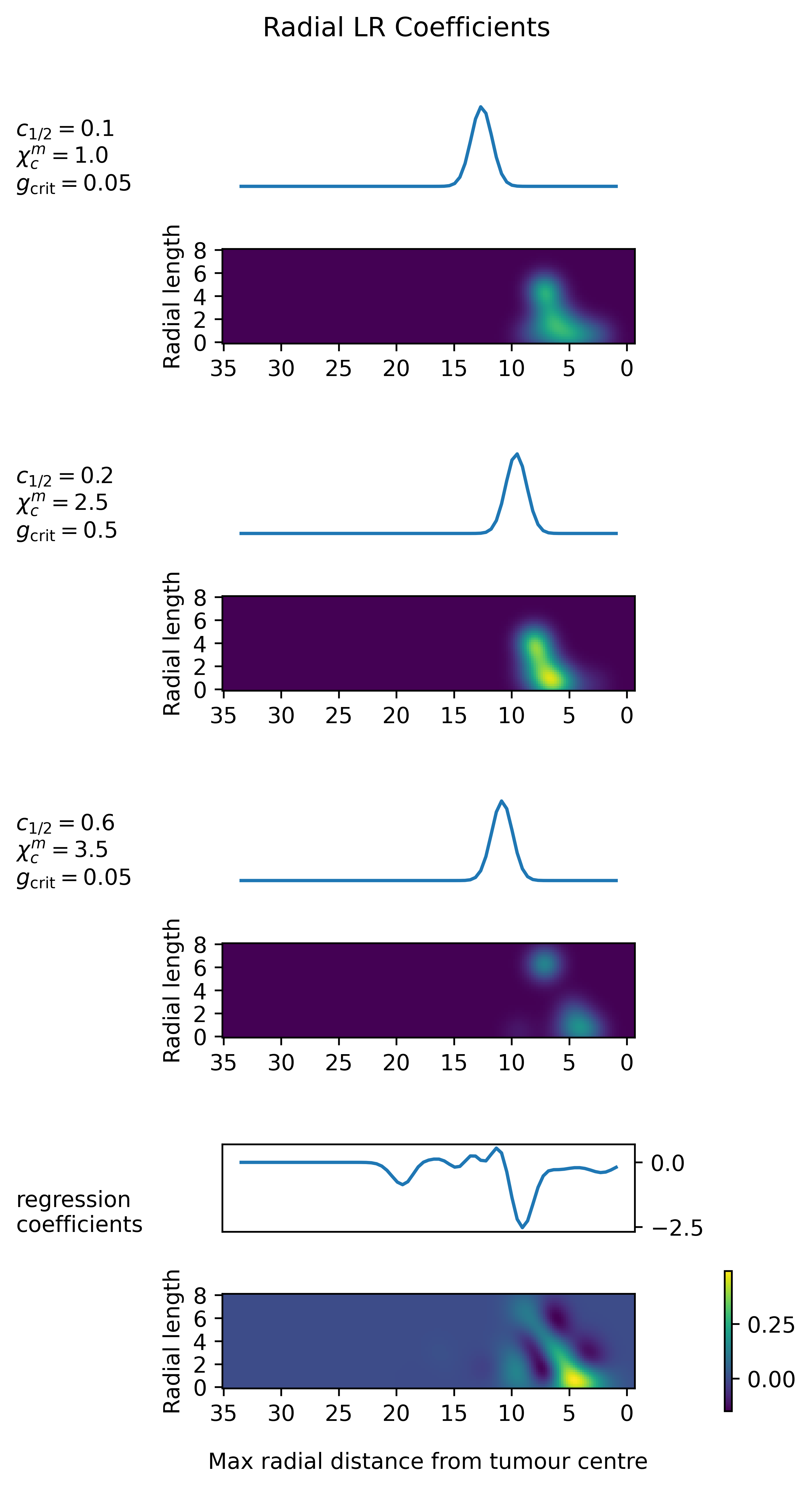}
    \caption{Sample persistence images $\radIm$ for the three example cases at $t = 350$. 1st row: Equilibrium case. 2nd row: Elimination case. 3rd row: Escape case. Last row: Logistic regression coefficients. The features in the persistence image represent separate outgrowths of the tumour. For a single outgrowth, birth ($x$-axis in the image) occurs at the furthest cell from the centre of the outgrowth, which we call the maximum radial distance of the feature. The persistence ($y$-axis) represents the length of its protrusion from the main body of the tumour in terms of the difference in radius. For example, a small outgrowth on a large tumour will have a high birth radius and low persistence and appear as a spot in the lower left of the persistence image. A large outgrowth on a small tumour will have a smaller birth distance and higher persistence, and it will appear in the upper right of the persistence image. The graph above each image records the infinite features which arise from completely disconnected components of the tumour. The graph will have a bump at the maximum radius of each component of the tumour. } 
    \label{fig:radial-results}
\end{figure}

\paragraph{Radial components --\boldmath$\radIm$}
The logistic regression coefficients and sample persistence images from the radial filtration are shown in Fig.~\ref{fig:radial-results}. 


The logistic regression coefficients show that infinite features are most prominent in the non-niche class, which is already seen in results from $\RipsIm_0(T)$ at $r = \varepsilon=0.7$. What the infinite features show beyond $\RipsIm_0(T)$ is the distribution of radii for relevant features. The logistic regression coefficients give a negative weight to almost all infinite features, but particularly those of radius around 10 and 20, with a slight positive weight to features of radius 12-15. A fragmented tumour would show a multi-modal curve of infinite features in $\radIm$ with a larger total area under the curve, so the overall negative logistic regression coefficients may be due to the prevalence of infinite features of $\radIm$. The positive regression coefficients may indicate that a main component radius around 12-15 might override other indicators. 

The finite logistic regression coefficients show a similar dependence on scale. 
Although the positive coefficients outweigh the negative coefficients, indicating that outgrowths in general are more strongly correlated with the niche case, there are regions that show the outgrowth pattern of non-niche cases. In the niche case, a tumour of radius 5 begins showing outgrowths, the length of which increases linearly as the tumour increases in radius, up to a tumour radius of 10 with a radial length of 6-8, long enough to reach a blood vessel. A higher radius, with low persistence, is associated with the non-niche class; from inspection, likely equilibrium. There is also an association of high persistence with a smaller radius to the non-niche class. We suspect this is due to the fragmentation that occurs in the elimination case, as with the infinite features.

The three featured examples all consist of a single tumour component, so the infinite feature curve has a single mode, centred on the maximum radius of the tumour. Our example curves show the largest radius in the equilibrium case and the lowest radius in the elimination case. The finite portions of $\radIm$ reveal strong boundary tortuosity in the elimination case, with a similar, but less dramatic, pattern for the equilibrium case.  
The escape case has fewer outgrowths, but significantly higher persistence of outgrowths. 



We note that the coefficients shown in Figure~\ref{fig:radial-results} apply only to timestep 350, so the values may change as the simulation progresses. If the values do vary between timesteps, we may 
need relatively precise normalisation of time and distances in future applications.



\subsection{Dynamic methods}\label{sec:time_results}

\begin{figure}[ht]
\includegraphics[width=0.105\textwidth]{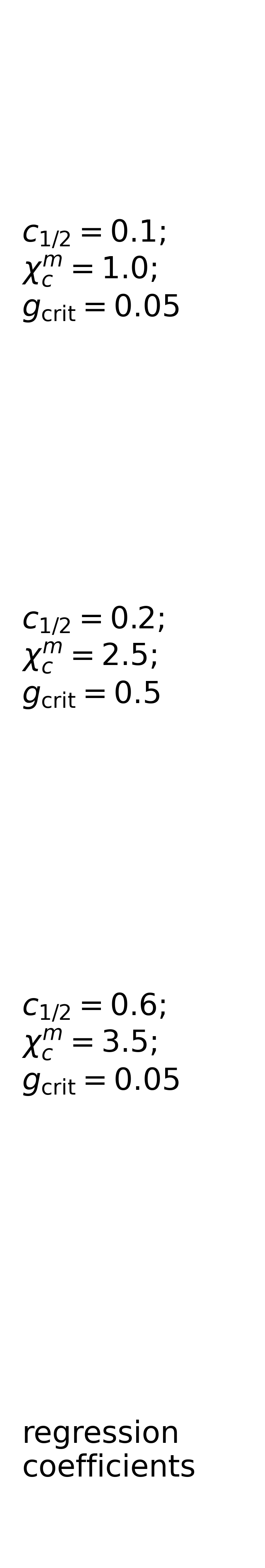}
\includegraphics[width=0.75\textwidth]{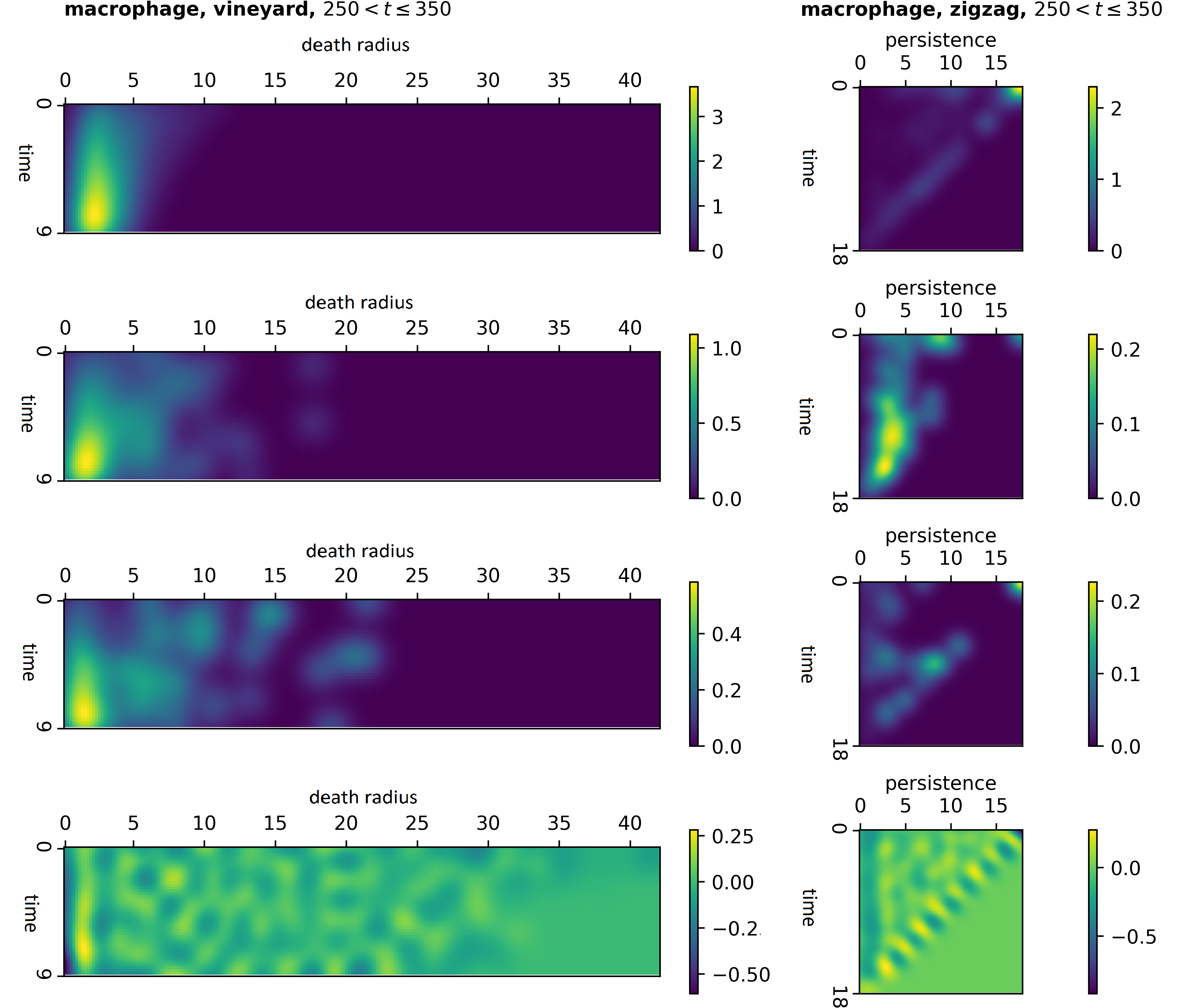}
\caption{{\bf Persistent images of selected techniques.}
The first three rows depict features extracted from the equilibrium, elimination, and escape simulations shown in Fig~\ref{fig:params}. For each, $\vinIm$ and $\zzIm$ are computed using 10 time-steps $250-340$. 1st row: Equilibrium case. 2nd row: Elimination case. 3rd row: Escape case. Last row: Regression coefficients shows the regions of the image which are most important for distinguishing whether or not the simulation would go on to produce a perivascular niche at time $t=500$.}
\label{fig:images_and_lr_time}
\end{figure}

The logistic regression coefficients and sample persistence images from the zigzag and vineyard filtrations of macrophage point clouds over the earliest time window, $250 \leq t \leq 350$ hours, are shown in Fig.~ \ref{fig:images_and_lr_time}. 

\setlength{\belowcaptionskip}{-10pt}

\subsubsection{Vineyards provide slight advantage over \boldmath$\RipsIm_0(M)$}

Since we envision real-world data to have a limited observational time window, we used a fairly short sequence. At a given time $t$, we compute the Vietoris-Rips persistence diagrams from 10 snapshots of the cell locations at times $t-90, t-80,  \dots, t-10, $ and $t$. We stack the Vietoris-Rips persistence diagrams and examine the variation of $\PH_0(\operatorname{VR}(X_t))$ as $t$ evolves.


Each trajectory in the image approximates a ``vine",  tracking the time evolution of a cluster of macrophages, from when the first macrophage emerges from a blood vessel, through changes in composition, scale, spacing, and phenotype, until it merges into a larger cluster or re-enters the blood vessel. 


Inspection of the persistence images $\vinIm$ reveals clear differences in the 
magnitude and patterns of persistence among the three qualitative behaviours (see Fig. \ref{fig:images_and_lr_time}). In all three cases, as with $\RipsIm_0(M)$, most of the relevant features occur at radius $0-2$. 
In the equilibrium example (top row), macrophages spaced 1-2 cell diameters apart appear to be the modal type of configuration, while the middle (elimination) case has overall fewer low-persistence features, with a higher proportion of large-scale features. These larger-scale features represent more isolated macrophage clusters. This pattern is even more pronounced in the bottom (pre-escape) case, with fewer clusters overall, and a higher percentage isolated from other clusters. 

The noisy pattern in the regression coefficients is difficult to interpret, as the diagrams are more complex, and the spotting in the coefficients may indicate over-fitting. However, the pattern generalises the logistic regression coefficients $\RipsIm_0(M)$, with a clear weighting on packed features ($0-1$ diameters) to identify non-niche cases, small-scale features ($1-2$ diam.) identifying the niche class, and larger features ($>2$ diam.) trending toward the non-niche class, possibly because of the niche class having fewer features overall. Interestingly, the niche coefficients are highest over timestep 8 ($t=340$), rather than 9 $(t=350)$. 
As the accuracy is comparable to $\RipsIm_0(M)$ until later time steps, we may not have captured some critical dynamics. For example, it appears in our sample images that the length and average isolation of vines may differ between the classes, but this cannot be captured in this vectorisation. Some alternatives are discussed in Section~\ref{sec:future}.


\subsubsection{Zigzags highlight enduring features}
Here each $K^t_\varepsilon$ is the static Vietoris-Rips complex at a fixed radius $\varepsilon$, and we use a variable range of ten time steps $(T-100,T-90,T-80,\dots,T-10)$ from an interval of length 90, ending at $T \in \{250, 300, 350, 400, 450, 500\}$. Along with an intersection complex between each consecutive time step, this gives a total sequence of length 19. Between time $t$ and $t+10$, we typically observe an optimal amount of change: noticeable evolution, but close enough in time to see continuity. 

To construct the intersection complex, we utilise the fixed numbering of cells in the simulation to accurately match cells between time steps. This is impossible in real data; for future work, we suggest an estimation methodology via optimal transport matching for real data (see Section~\ref{sec:real-data}). We compute the zigzag barcode for dimension 0 using \texttt{BATS} \cite{Carlsson:zigzag-computational}. A bar with birth time $b$ and death time $d$ indicates a topological feature that persists through the zigzag filtration between the $b$-th and $d$-th complexes. 

\paragraph{Linking radius}
The uniform radius $\varepsilon=2$ is selected from a coarse sweep, in the same manner as for the radial filtration in Fig.~\ref{fig:eps_rad}. However, the optimal $\varepsilon$ is significantly larger for the macrophages than the tumour cells, suggesting that macrophages with up to three cells between them should be considered as a functional ``cluster". 
%
%
As macrophages, unlike tumours, move independently, which suggests that 
whether or not macrophages in close proximity remain in proximity, working in concert, may be relevant to future outcomes. 

\paragraph{Macrophage zigzags --$\mathbf{\zzIm}$.}
 We can see clear differences between the $\zzIm$ in the three example cases (Fig.~\ref{fig:images_and_lr_time}), with the equilibrium simulation (top) exhibiting by far the highest number of macrophage clusters, with over twice as many new macrophage clusters forming during the window that persist through the end. In contrast, the middle image, representing the elimination case before elimination has occurred, shows less overall activity, with the frequent appearance of new features and most features persisting for only a few time steps. The escape case in the bottom $\zzIm$ shows a much higher proportion of features along the diagonal than the elimination case, with about the same overall activity level. This suggests that the number of macrophage clusters is comparable, but clusters are more static over time.
 
We examine the logistic regression coefficients to explore the effect of static macrophage clusters. The most distinctive feature of the zigzag regression coefficient array is the diagonal, which represents features that persist through to the end of the zigzag. The diagonal also has a distinct alternating pattern, with negative coefficients at even birth times and positive coefficients at odd ones. As alternating complexes are intersections of the two timesteps on either side, very different behaviour occurs at even (original data-based) and odd (intersection) filtration steps, as shown in Figure~\ref{fig:zigzag}. Connected components born from the appearance of new macrophages will appear in the even (data) steps, and new connected components arising from the division of existing connected components will appear in the odd (intersection) step. 

The results clearly show a pattern that is not readily apparent from inspecting the images: the niche case is correlated with the division or rearrangement of macrophage clusters from one timestep to the next that remain stable over time, while the non-niche case is correlated with the appearance of new macrophages which remain in the simulation, isolated from other macrophages, for the duration of the time window (see Fig. \ref{fig:images_and_lr_time}). Additionally, features that have persistence 1, that is, clusters that appear for a single time step, are associated with the non-niche class. Based on our example, this appears to be characteristic of the elimination case.



\subsection{Prediction accuracy}
\label{sec:accuracies}
\begin{figure}[ht]
\includegraphics[width=\textwidth]{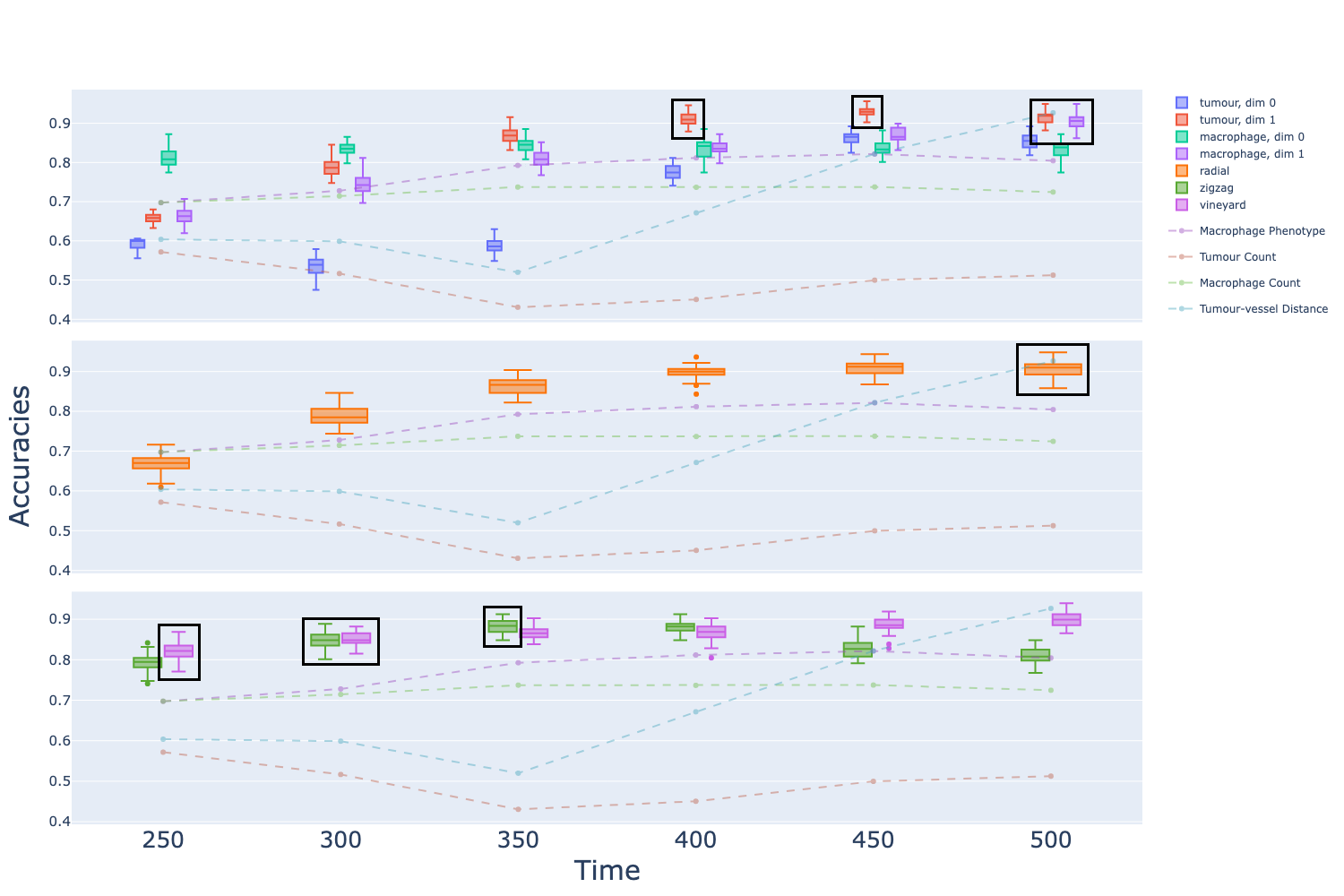}
\caption{{\bf Accuracy against time.} Accuracy scores in predicting the formation of at least one perivascular niche at $500$ hours. Scores are given for the different filtrations based on cell locations at time steps $t=250,300,350,400,450,$ and $500$ for the static filtrations, and for each value of $t$, the range $t-100,t-90\dots, t-20,t-10$ for the dynamic. We compare the four topological methods to various benchmarks associated with simple statistics (dotted lines). The highest scoring methods at each time step are boxed. Top: Vietoris-Rips filtrations. Centre: The tumour radial filtration. Bottom: Zigzag and vineyard filtrations.}
\label{fig:summary}
\end{figure}

The percentage of test samples that are correctly predicted to have a perivascular niche or not at time $t = 500$ is given by the accuracy score. 
We compute the prediction accuracy scores for each method (Fig.~\ref{fig:summary}). 

\paragraph{Comparison to benchmarks. }
In addition to the topological methods, we also implemented classifiers trained on four benchmark statistics: macrophage phenotype ratio, tumour cell count, macrophage count, and minimum tumour-vessel distance. We selected these statistics to provide a competitive benchmark relating our work to other methods trained on the same data. Macrophage phenotype is often used for classification, usually with a choice of threshold; here we have used the continuous phenotype ratio values and let the classifier set the decision boundary optimally. 
Tumour cell count, a proxy for tumour size, will obviously be much lower in tumour elimination cases, giving this measurement an edge in detecting elimination behaviour. Tumour-vessel distance should naturally perform very well, given our definition of a perivascular niche (i.e. the presence of at least 10 tumour cells in close proximity to a blood vessel). High accuracy at time step 500 is therefore almost a certainty, with tumour-vessel distance at late time-steps expected to correlate accordingly.

All of the topological methods (see Figure \ref{fig:summary}) consistently outperform tumour cell counts, which do not accurately predict perivascular niches at any time, and most outperform tumour-vessel distance, which has fairly low accuracy until late time steps. 


As benchmarks, we note that the macrophage count and phenotype ratio, which do not involve tumour-vessel information, achieve accuracies of at least 70\%, with the phenotype ratio outperforming the macrophage count. All of the topological techniques achieve similar or better accuracies, at much earlier timesteps, except for $\RipsIm_0(T)$.
We hypothesize that like the tumour cell count, $\RipsIm_0(T)$ has low accuracy due to its difficulty distinguishing the equilibrium and escape cases.

\paragraph{Highest accuracy uses tumour shape. } Aside from tumour-vessel distance at time $500$ (measurement of which is similar to testing class membership directly), tumour-based PH, including $\RipsIm_1(T)$ and $\radIm$, provide the highest accuracy results in our experiment. $\RipsIm_1(T)$ and $\radIm$ achieve above $90\%$ accuracy for time steps $400 \leq t \leq 500$, in most cases before perivascular niches appear.

$\RipsIm_1(T)$ and $\radIm$ track the non-compactness of the tumour, including (resp.) holes in the interior and outgrowths from the tumour. 
The high accuracy scores confirm classic intuition that tumour components and their placement, as well as the regularity of the tumour boundary, including extended outgrowths from the main tumour body, are practical predictors that a perivascular niche is about to form. See Sections \ref{sec:VR_results} and \ref{sec:rad_results} for details.
%
Between the two measures, $\RipsIm_1(T)$ performs slightly better than $\radIm$, although they encode 
different shape information about the tumour. Computationally, $\radIm$ requires only $\operatorname{PH}_0$, which can be more efficient for large data sets than $\RipsIm_1.$

We emphasise that none of the topological summaries utilise the location of blood vessels or tumour-vessel distance; rather, the measurements are intrinsic to the spatial patterns of macrophages and tumour cells. The closest proxy is the radial filtration $\radIm$, which detects distances between tumour cells and the original tumour centre at the midpoint of the simulation. As the blood vessels are located at large radius, there should be high (inverse) correlation between radial birth values and tumour-vessel distance. Nevertheless, $\radIm$ outperforms tumour-vessel distance until the final timestep.
%

\paragraph{Dynamic methods enable earlier detection.}
Among all techniques analysed, macrophage-based homology, including multi-scale clusters $\RipsIm_0(M)$, zigzag clusters $\zzIm$, and vineyards of Rips diagrams $\vinIm$ achieve the highest accuracy in the first half of the time frame studied, at values $250 \leq t \leq 350.$ By including dynamic topological information in $\vinIm$ and $\zzIm$, we are able to achieve a comparable level of accuracy up to 50 time steps earlier than the best-performing static methods. All three macrophage cluster-based methods achieve $80\%$ accuracy 100 time steps earlier than phenotype ratio and 200 time steps earlier than tumour-vessel distance - almost half the lifetime of the simulation. 

The strength of these results suggest that given fixed tumour parameters, perivascular niches can be predicted using the arrangements of macrophages alone - above and beyond the number of macrophages, their clustering patterns reflect differences in critical threshold and chemotactic sensitivity. 

\section{Conclusion and Future Work}

In order to better understand the qualitative behaviour of tumour-immune interactions, we studied a series of topological summaries which represent the shape of a point cloud as a persistence image. We tested these methods on an agent-based model, to try to predict which simulations would go on to form a perivascular niche, and which would not. 


%
We compared the corresponding topological vectorisations to alternative, benchmarking features (e.g., tumour size, macrophage phenotype, and distance from tumour to blood vessel).
We found persistence images, encoding the geometry and topology of the point clouds of tumour and macrophage cells, performed exceptionally well at predicting the future development of perivascular niches. 
%
%

At early stages, starting at the halfway point of the simulation $t=250$, topological summaries based on the macrophage point cloud were able to predict perivascular niches with $80\%$ accuracy, matching the accuracy of macrophage phenotype ratio over 100 time steps earlier. Dynamic methods $\vinIm$ and $\zzIm$ allowed for slightly earlier classification than static macrophage multi-scale clusters, by capturing the topological evolution of the point cloud of macrophages over the preceding 90 hours.

At time steps $400$, topological summaries based on holes and tortuosity in the tumour point cloud achieve $90\%$ accuracy in the prediction of perivascular niche formation.
Based on their performance, we can confirm intuition that 1-dimensional features, or holes, inside the tumour point cloud and irregularity of the tumour boundary are highly predictive. The logistic regression coefficients contained even richer detail on the geometric features relevant for each class, showing that it is larger holes, and tumour outgrowths of a particular length, that are most indicative of potential for a perivascular niche, with macrophage accumulation around the boundary of a tumour indicating a lower likelihood of niche formation.


These results lead naturally to future research directions, including the application of the topological techniques to real data, as described in Section \ref{sec:real-data}, and extension of the methods 
%
%
to more sophisticated topological summaries, as described in Section \ref{sec:future}.

\subsection{Adapting the methods to biological data}\label{sec:real-data}


The true strengths and limitations of any method only become apparent when they are evaluated in real-world applications.
We believe, nonetheless, that the pipelines we have developed can be easily applied to other synthetic agent-based model data as well as adapted to other spatio-temporal experimental/clinical data . 
While most of our topological constructions could be applied to real data sets without modification, the zigzag and radial complexes require some adaptation. The radial complex requires a choice of basepoint. In our analysis, we fixed the basepoint at the centre of the simulation, but in imaging data from cancer, the tumour's centre of mass could be used similar to radial filtration applied to tumour vascular networks \cite{stolz2022}. Further, both the radial and zigzag require a fixed linkage radius, which depends on the size and density of the cells. 

To construct the intersection complex for the zigzag complex, we used information from the simulations (the identities of each autonomous cell) that would only be available in live-tracking data, but not other snapshot capture experiments. A point is present in the intersection complex if the corresponding cell is present in both timesteps, even if it has moved significantly. Similarly, higher simplices are present if they are present in both $K^t_\varepsilon$ and $K^{(t+10)}_\varepsilon$, that is, if the cells remain in (or return to) proximity within the interval. Since real data  will typically not contain individual cell trajectories of cells, it may not be possible to determine which simplices are new, and which have moved. We propose that cell identity would be best approximated by continuity, finding $M$ which minimises $ \inf_M \left(\sum_{(m_1,m_2)\in M}||m_1-m_2||^2\right) ,$
where $M = \{(m_1,m_2)\}$ is a matching of $M_1 = M_t \cup V$ and $M_2 = M_{t+10}\cup V$ which gives a bijection between $M_1$ and $M_2$, but with arbitrary multiplicity on blood vessel set $V$.
Implementation of this method on our synthetic data yielded similar results in preliminary testing, but we postpone further development of this approach to future research. 

Finally, data sets of different sizes and/or scales would require normalization of the persistence images. We have chosen to vectorise our diagrams into persistence images in part for the ease of normalization, as images can be cropped, padded, or scaled easily. As the local behaviour of cells may be comparable for areas (and tumours) of any size, we recommend normalising the images to the same birth axis.

\subsection{Further methods and analysis}
\label{sec:future}
As several of our methods individually outperformed the benchmarks, with different features and strengths, we naturally would consider combining several vectorisations in order to increase predictive accuracy. Our results suggest that a macrophage-based clustering, such as $\vinIm$ (or $\RipsIm_0(M)$ where fine time steps are not available), combined with tumour $\RipsIm_1(T)$ or $\radIm$, would be particularly powerful when the stage of development is not known.

It is also natural to consider 
combining 
into a multiparameter persistence module. Multipersistence has already been applied to study spatial patterns of immune cells in tumours \cite{MPH-tumour}. The radial filtration, for example, could be generalised to avoid choosing a linkage radius, so that the complex is simultaneously filtered on distance from the centre and scale, as in the Vietoris-Rips complex.
We may also bypass the same issue of having to choose an $\varepsilon$ in the zigzag filtration by studying spatio-temporal data with an interlevel-Rips trifiltration \cite{kim2021spatiotemporal}. To our knowledge, no software exists for such analyses.

A second obvious direction would be the extension to 3-dimensional models and data. Although we developed these techniques for a two-dimensional spatial model, the methods described can also be applied to 3-dimensional point clouds with little modification. Indeed, many of the works cited analyse 3-d data with similar techniques, and we expect our results to extend naturally. In 3-d data it is possible to identify higher homology groups, representing loops and enclosed voids in the 3-d filtrations. These higher order features may provide additional classifying power. 

%
%

Finally, we note that persistence diagrams can be vectorised in several different ways, and persistence images are one choice among many others such as \textit{persistence landscapes} \cite{bubenik:landscape} and 
 \textit{CROCKER plots} (used in \cite{Topaz2015}). Persistence flamelets have also been suggested as a robust summary for time-varying data; they could serve as a reasonable substitute for our vineyard vectorisation \cite{flamelets}. While we anticipate that persistence images are likely to offer the most robust results for this type of analysis, this remains to be tested. 

\section*{Supporting information}
Our code and data can be found at \url{https://github.com/Pyxidatol-C/ABM-TDA}.

\section*{Acknowledgments}
J.Y., H.F., H.A.H. and G.G. are supported by the Royal Society grant RGF{\textbackslash}EA{\textbackslash}201074. 
J.A.B. is supported by Cancer Research UK grant number CTRQQR-2021\textbackslash100002, through the Cancer Research UK Oxford Centre.
H.A.H. gratefully acknowledges funding from EPSRC EP/R018472/1, EP/R005125/1 and EP/T001968/1, the Royal Society RGF{\textbackslash}EA{\textbackslash}201074 and UF150238. 
H.A.H. and H.M.B. acknowledge funding from the Emerson Collective. 
H.M.B. acknowledges funding from the Ludwig Institute for Cancer Research (Oxford Branch).
I.Y.\ and H.M.B.\ acknowledge funding from the Mark Foundation.
G.G. also acknowledges funding from National Science Foundation MSPRF 2202895.
For the purpose of Open Access, the authors have applied a CC BY public copyright licence to any Author Accepted Manuscript (AAM) version arising from this submission.


Many thanks to Robert A.\ McDonald and David Beers for informative meetings; to Wilf Offred and Mark D.\ Williams for spending a productive summer full of laughter together.


\printbibliography

@article{1PH:stability,
    author = {Ghrist, Robert},
    year = {2008},
    month = {02},
    pages = {},
    title = {Barcodes: The persistent topology of data},
    volume = {45},
    journal = {BULLETIN (New Series) OF THE AMERICAN MATHEMATICAL SOCIETY},
    doi = {10.1090/S0273-0979-07-01191-3}
}

@article{stolz2022,
  title={Multiscale topology characterizes dynamic tumor vascular networks},
  author={Stolz, Bernadette J and Kaeppler, Jakob and Markelc, Bostjan and Braun, Franziska and Lipsmeier, Florian and Muschel, Ruth J and Byrne, Helen M and Harrington, Heather A},
  journal={Science Advances},
  volume={8},
  number={23},
  year={2022},
  publisher={American Association for the Advancement of Science},
  doi = {10.1126/sciadv.abm2456}

}

@article{kim2021spatiotemporal,
  title={Spatiotemporal persistent homology for dynamic metric spaces},
  author={Kim, Woojin and M{\'e}moli, Facundo},
  journal={Discrete \& Computational Geometry},
  volume={66},
  pages={831--875},
  year={2021},
  publisher={Springer},
  url = {https://doi.org/10.1007/s00454-019-00168-w},
  doi = {10.1007/s00454-019-00168-w}
}

@article{bubenik2020persistent,
  title={Persistent homology detects curvature},
  author={Bubenik, Peter and Hull, Michael and Patel, Dhruv and Whittle, Benjamin},
  journal={Inverse Problems},
  volume={36},
  number={2},
  year={2020},
  publisher={IOP Publishing},
  doi = {10.1088/1361-6420/ab4ac0}
}

@phdthesis{stolz-pretzer2019a,
  publisher = {University of Oxford},
  school = {University of Oxford},
  title = {Global and local persistent homology for the shape and classification of biological data},
  author = {Stolz-Pretzer, B},
  year = {2019},
}

@article{kanari,
  title={A topological representation of branching neuronal morphologies},
  author={Kanari, Lida and D{\l}otko, Pawe{\l} and Scolamiero, Martina and Levi, Ran and Shillcock, Julian and Hess, Kathryn and Markram, Henry},
  journal={Neuroinformatics},
  volume={16},
  number={1},
  pages={3--13},
  year={2018},
  publisher={Springer},
  doi = {https://doi.org/10.1007/s12021-017-9341-1}
}

@misc{ali2022survey,
      title={A Survey of Vectorization Methods in Topological Data Analysis}, 
      author={Dashti Ali and Aras Asaad and Maria-Jose Jimenez and Vidit Nanda and Eduardo Paluzo-Hidalgo and Manuel Soriano-Trigueros},
      year={2022},
      eprint={2212.09703},
      archivePrefix={arXiv},
      primaryClass={math.AT}
}

@article{carlsson2009topology,
  title={Topology and data},
  author={Carlsson, Gunnar},
  journal={Bulletin of the American Mathematical Society},
  volume={46},
  number={2},
  pages={255--308},
  year={2009},
  doi = {https://doi.org/10.1090/S0273-0979-09-01249-X}
}

@article{ghrist2008barcodes,
  title={Barcodes: the persistent topology of data},
  author={Ghrist, Robert},
  journal={Bulletin of the American Mathematical Society},
  volume={45},
  number={1},
  pages={61--75},
  year={2008},
  url = {https://www.ams.org/journals/bull/2008-45-01/S0273-0979-07-01191-3/S0273-0979-07-01191-3.pdf},
  doi = {https://doi.org/10.1090/S0273-0979-07-01191-3}
}

@article{MPH-tumour,
    author = {Oliver Vipond  and Joshua A. Bull  and Philip S. Macklin  and Ulrike Tillmann  and Christopher W. Pugh  and Helen M. Byrne  and Heather A. Harrington },
    title = {Multiparameter persistent homology landscapes identify immune cell spatial patterns in tumors},
    journal = {Proceedings of the National Academy of Sciences},
    volume = {118},
    number = {41},
    pages = {e2102166118},
    year = {2021},
    doi = {10.1073/pnas.2102166118},
    URL = {https://www.pnas.org/doi/abs/10.1073/pnas.2102166118}
}

@misc{Rob:zigzag-coral,
  doi = {10.48550/ARXIV.2209.08974},
  url = {https://arxiv.org/abs/2209.08974},
  author = {McDonald, Robert A. and Neuhausler, Rosanna and Robinson, Martin and Larsen, Laurel G. and Harrington, Heather A. and Bruna, Maria},
  title = {Topological descriptors for coral reef resilience using a stochastic spatial model},
  publisher = {arXiv},
  year = {2022},
  copyright = {Creative Commons Attribution 4.0 International}
}

@article {Josh:ABM,
	author = {Bull, Joshua A. and Byrne, Helen M.},
	title = {Quantification of spatial and phenotypic heterogeneity in an agent-based model of tumour-macrophage interactions},
	year = {2023},
	doi = {10.1371/journal.pcbi.1010994},
	journal = {PLoS Computational Biology},
    volume = {19},
    issue = {3},
    pages = {e1010994}
}

@article {chaste,
	author = {Cooper, Fergus R. and Baker, Ruth E. and Bernabeu, Miguel O. and Bordas, Rafel and Bowler, Louise and Bueno-Orovio, Alfonso and Byrne, Helen M. and Carapella, Valentina and Cardone-Noott, Louie and Cooper, Jonathan and Dutta, Sara and Evans, Benjamin D. and Fletcher, Alexander G. and Grogan, James A. and Guo, Wenxian and Harvey, Daniel G. and Hendrix, Maurice and Kay, David and Kursawe, Jochen and Maini, Philip K. and McMillan, Beth and Mirams, Gary R. and Osborne, James M. and Pathmanathan, Pras and Pitt-Francis, Joe M. and Robinson, Martin and Rodriguez, Blanca and Spiteri, Raymond J. and Gavaghan, David J.},
	title = {Chaste: Cancer, Heart and Soft Tissue Environment},
	year = {2020},
	doi = {10.21105/joss.01848},
	journal = {The Journal of Open Source Software}
}

@article{Vipond:MPH-landscape,
  author  = {Oliver Vipond},
  title   = {Multiparameter Persistence Landscapes},
  journal = {Journal of Machine Learning Research},
  year    = {2020},
  volume  = {21},
  number  = {61},
  pages   = {1--38},
  url     = {http://jmlr.org/papers/v21/19-054.html}
}

@misc{Carlsson:zigzag-computational,
  doi = {10.48550/ARXIV.1911.10693},
  url = {https://arxiv.org/abs/1911.10693},
  author = {Carlsson, Gunnar and Dwaraknath, Anjan and Nelson, Bradley J.},
  title = {Persistent and Zigzag Homology: A Matrix Factorization Viewpoint},
  publisher = {arXiv},
  year = {2019},
  copyright = {arXiv.org perpetual, non-exclusive license}
}

@article{Carlsson:zigzag,
  title={Zigzag persistence},
  author={Carlsson, Gunnar E. and De Silva, Vin},
  journal={Foundations of computational mathematics},
  volume={10},
  pages={367--405},
  year={2010},
  publisher={Springer},
  url = {https://doi.org/10.1007/s10208-010-9066-0},
  doi = {10.1007/s10208-010-9066-0}
}

@ARTICLE{colorectal,
  title     = "Elevated {CD163+/CD68+} ratio at tumor invasive front is closely
               associated with aggressive phenotype and poor prognosis in
               colorectal cancer",
  author    = "Yang, Chaogang and Wei, Chen and Wang, Shuyi and Shi, Dongdong
               and Zhang, Chunxiao and Lin, Xiaobin and Dou, Rongzhang and
               Xiong, Bin",
  journal   = "Int. J. Biol. Sci.",
  publisher = "Ivyspring International Publisher",
  volume    =  15,
  number    =  5,
  pages     = "984--998",
  month     =  mar,
  year      =  2019,
  language  = "en",
  url = "https://www.ncbi.nlm.nih.gov/pmc/articles/PMC6535793/",
  doi = "10.7150/ijbs.29836"
}

@ARTICLE{ovarian,
  title     = "A high {M1/M2} ratio of tumor-associated macrophages is
               associated with extended survival in ovarian cancer patients",
  author    = "Zhang, Meiying and He, Yifeng and Sun, Xiangjun and Li, Qing and
               Wang, Wenjing and Zhao, Aimin and Di, Wen",
  journal   = "J. Ovarian Res.",
  publisher = "Springer Science and Business Media LLC",
  volume    =  7,
  number    =  1,
  pages     = "19",
  month     =  feb,
  year      =  2014,
  language  = "en",
  url = "https://doi.org/10.1186/1757-2215-7-19"
}

@misc{covid-vineyard,
  doi = {10.48550/ARXIV.2107.09188},
  
  url = {https://arxiv.org/abs/2107.09188},
  
  author = {Hickok, Abigail and Needell, Deanna and Porter, Mason A.},

  title = {Analysis of Spatial and Spatiotemporal Anomalies Using Persistent Homology: Case Studies with COVID-19 Data},
  publisher = {arXiv},
  year = {2021},
  copyright = {arXiv.org perpetual, non-exclusive license}
}

@article{immune-tumour,
  title={Roles of the immune system in cancer: from tumor initiation to metastatic progression},
  author={Gonzalez, Hugo and Hagerling, Catharina and Werb, Zena},
  journal={Genes \& development},
  volume={32},
  number={19-20},
  pages={1267--1284},
  year={2018},
  publisher={Cold Spring Harbor Lab},
  doi = {10.1101/gad.314617.118}
}

@inproceedings{prognostic,
  title={Prognostic significance of tumor-associated macrophages: Past, present and future},
  author={Cortese, Nina and Carriero, Roberta and Laghi, Luigi and Mantovani, Alberto and Marchesi, Federica},
  booktitle={Seminars in immunology},
  volume={48},
  year={2020},
  organization={Elsevier},
  url = {https://doi.org/10.1016/j.smim.2020.101408},
  doi = {10.1016/j.smim.2020.101408}
}

@article{immunotherapy,
  title={Harnessing the antitumor potential of macrophages for cancer immunotherapy},
  author={Long, Kristen B and Beatty, Gregory L},
  journal={Oncoimmunology},
  volume={2},
  number={12},
  year={2013},
  publisher={Taylor \& Francis},
  url = {https://doi.org/10.4161/onci.26860},
  doi = {10.4161/onci.26860}
}

@article{zigzag-stability,
	doi = {10.2140/agt.2018.18.3133},
  
	url = {https://doi.org/10.2140%2Fagt.2018.18.3133},
  
	year = 2018,
	month = {10},
  
	publisher = {Mathematical Sciences Publishers},
  
	volume = {18},
  
	number = {6},
  
	pages = {3133--3204},
  
	author = {Magnus Botnan and Michael Lesnick},
  
	title = {Algebraic stability of zigzag persistence modules},
  
	journal = {Algebraic {\&} Geometric Topology}
}

@misc{zigzag-decomposition,
  doi = {10.48550/ARXIV.1507.01899},
  
  url = {https://arxiv.org/abs/1507.01899},
  
  author = {Botnan, Magnus Bakke},
  
  title = {Interval Decomposition of Infinite Zigzag Persistence Modules},
  
  publisher = {arXiv},
  
  year = {2015},
  
  copyright = {arXiv.org perpetual, non-exclusive license}
}

@article{persistence-images,
  author  = {Henry Adams and Tegan Emerson and Michael Kirby and Rachel Neville and Chris Peterson and Patrick Shipman and Sofya Chepushtanova and Eric Hanson and Francis Motta and Lori Ziegelmeier},
  title   = {Persistence Images: A Stable Vector Representation of Persistent Homology},
  journal = {Journal of Machine Learning Research},
  year    = {2017},
  volume  = {18},
  number  = {8},
  pages   = {1--35},
  url     = {http://jmlr.org/papers/v18/16-337.html}
}

@inproceedings{vineyard,
author = {Cohen-Steiner, David and Edelsbrunner, Herbert and Morozov, Dmitriy},
title = {Vines and Vineyards by Updating Persistence in Linear Time},
year = {2006},
isbn = {1595933409},
publisher = {Association for Computing Machinery},
address = {New York, NY, USA},
url = {https://doi.org/10.1145/1137856.1137877},
doi = {10.1145/1137856.1137877},
booktitle = {Proceedings of the Twenty-Second Annual Symposium on Computational Geometry},
pages = {119–126},
numpages = {8},
location = {Sedona, Arizona, USA},
series = {SCG '06}
}

@article{model1,
  doi = {10.1142/s0218202599000270},
  url = {https://doi.org/10.1142/s0218202599000270},
  year = {1999},
  month = jun,
  publisher = {World Scientific Pub Co Pte Lt},
  volume = {09},
  number = {04},
  pages = {513--539},
  author = {Markus R. Owen and Jonathan A. Sherratt},
  title = {Mathematical modelling of macrophage dynamics in tumours},
  journal = {Mathematical Models and Methods in Applied Sciences}
}

@article{model2,
author={Kelly, C. E.
and Leek, R. D.
and Byrne, H. M.
and Cox, S. M.
and Harris, A. L.
and Lewis, C. E.},
title={Modelling Macrophage Infiltration into Avascular Tumours},
journal={Journal of Theoretical Medicine},
year={1900},
month={1},
day={01},
publisher={Hindawi Publishing Corporation},
volume={4},
pages={190323},
issn={1748-670X},
doi={10.1080/10273660290015242},
url={https://doi.org/10.1080/10273660290015242}
}

@ARTICLE{model3,
  title     = "Mathematical modelling of the use of macrophages as vehicles for
               drug delivery to hypoxic tumour sites",
  author    = "Owen, Markus R and Byrne, Helen M and Lewis, Claire E",
  journal   = "J. Theor. Biol.",
  publisher = "Elsevier BV",
  volume    =  226,
  number    =  4,
  pages     = "377--391",
  month     =  feb,
  year      =  2004,
  language  = "en",
 url = "https://doi.org/10.1016/j.jtbi.2003.09.004",
 doi = "10.1016/j.jtbi.2003.09.004"
}

@ARTICLE{model4,

AUTHOR={Li, Xuefei and Jolly, Mohit Kumar and George, Jason T. and Pienta, Kenneth J. and Levine, Herbert},

TITLE={Computational Modeling of the Crosstalk Between Macrophage Polarization and Tumor Cell Plasticity in the Tumor Microenvironment},

JOURNAL={Frontiers in Oncology},

VOLUME={9},

YEAR={2019},

URL={https://www.frontiersin.org/articles/10.3389/fonc.2019.00010},

DOI={10.3389/fonc.2019.00010},

ISSN={2234-943X}
}

@article{model5,
title = {The re-polarisation of M2 and M1 macrophages and its role on cancer outcomes},
journal = {Journal of Theoretical Biology},
volume = {390},
pages = {23-39},
year = {2016},
issn = {0022-5193},
doi = {10.1016/j.jtbi.2015.10.034},
url = {https://www.sciencedirect.com/science/article/pii/S0022519315005330},
author = {Nicoline Y. {den Breems} and Raluca Eftimie}
}

@Article{model6,
author={El-Kenawi, Asmaa
and Gatenbee, Chandler
and Robertson-Tessi, Mark
and Bravo, Rafael
and Dhillon, Jasreman
and Balagurunathan, Yoganand
and Berglund, Anders
and Vishvakarma, Naveen
and Ibrahim-Hashim, Arig
and Choi, Jung
and Luddy, Kimberly
and Gatenby, Robert
and Pilon-Thomas, Shari
and Anderson, Alexander
and Ruffell, Brian
and Gillies, Robert},
title={Acidity promotes tumour progression by altering macrophage phenotype in prostate cancer},
journal={British Journal of Cancer},
year={2019},
month={10},
day={01},
volume={121},
number={7},
pages={556-566},
issn={1532-1827},
doi={10.1038/s41416-019-0542-2},
url={https://doi.org/10.1038/s41416-019-0542-2}
}

@article{Knutsdottir2014,
  doi = {10.1016/j.jtbi.2014.04.031},
  year = {2014},
  volume = {357},
  pages = {184-199},
  author = {H Kn{\'{u}}tsd{\'{o}}ttir and E P{\'{a}}lsson and L Edelstein-Keshet},
  title = {Mathematical model of macrophage-facilitated breast cancer cells
  invasion},
  journal = {Journal of Theoretical Biology}
}

@article{Knutsdottir2016,
  doi = {10.1039/C5IB00201J},
  year = {2016},
  volume = {8},
  issue = {1},
  pages = {104-119},
  author = {H Kn{\'{u}}tsd{\'{o}}ttir and J Condeelis and E P{\'{a}}lsson},
  title = {3-D individual cell based computational modeling of tumor
  cell–macrophage paracrine signaling mediated by EGF and CSF-1 gradients},
  journal = {Integrative Biology}
}

@article{Elitas2016,
  doi = {10.19070/2167-9118-SI05001},
  year = {2016},
  volume = {S5},
  issue = {001},
  pages = {1-8},
  author = {M Elitas and S Zeinali},
  title = {Modeling and Simulation of EGF-CSF-1 pathway to Investigate Glioma -
  Macrophage Interaction in Brain Tumors},
  journal = {International Journal of Cancer Studies \& Research (IJCR)}
}

@article{Eftimie2020,
  doi = {10.1016/j.mbs.2020.108325},
  year = {2020},
  volume = {322},
  issue = {March 2019},
  pages = {108325},
  author = {R Eftimie},
  title = {Investigation into the role of macrophages heterogeneity on solid
  tumour aggregations},
  journal = {Mathematical Biosciences}
}

@article{Topaz2015,
  doi = {10.1371/journal.pone.0126383},
  url = {https://doi.org/10.1371/journal.pone.0126383},
  year = {2015},
  month = may,
  publisher = {Public Library of Science ({PLoS})},
  volume = {10},
  number = {5},
  pages = {e0126383},
  author = {Chad M. Topaz and Lori Ziegelmeier and Tom Halverson},
  editor = {Bard Ermentrout},
  title = {Topological Data Analysis of Biological Aggregation Models},
  journal = {{PLOS} {ONE}}
}

@inproceedings{stability_Chazal,
author = {Chazal, Fr\'{e}d\'{e}ric and Cohen-Steiner, David and Glisse, Marc and Guibas, Leonidas J. and Oudot, Steve Y.},
title = {Proximity of Persistence Modules and Their Diagrams},
year = {2009},
isbn = {9781605585017},
publisher = {Association for Computing Machinery},
address = {New York, NY, USA},
url = {https://doi.org/10.1145/1542362.1542407},
doi = {10.1145/1542362.1542407},
booktitle = {Proceedings of the Twenty-Fifth Annual Symposium on Computational Geometry},
pages = {237–246},
numpages = {10},
location = {Aarhus, Denmark},
series = {SCG '09}
}

@article{persistence-diagram:stability,
  doi = {10.1007/s00454-006-1276-5},
  url = {https://doi.org/10.1007/s00454-006-1276-5},
  year = {2006},
  month = dec,
  publisher = {Springer Science and Business Media {LLC}},
  volume = {37},
  number = {1},
  pages = {103--120},
  author = {David Cohen-Steiner and Herbert Edelsbrunner and John Harer},
  title = {Stability of Persistence Diagrams},
  journal = {Discrete \& Computational Geometry}
}

@article{Bauer2021Ripser,
	Author = {Ulrich Bauer},
	Title = {Ripser: efficient computation of Vietoris-Rips persistence barcodes},
	Journal = {Journal of Applied and Computational Topology},
	Year = {2021},
	Publisher = {Springer},
	DOI = {10.1007/s41468-021-00071-5}
}

@article{bubenik:landscape,
  doi = {10.48550/ARXIV.1207.6437},
  
  url = {https://arxiv.org/abs/1207.6437},
  
  author = {Bubenik, Peter},
  
  title = {Statistical topological data analysis using persistence landscapes},
  
  publisher = {arXiv},
  
  year = {2012},
  
  copyright = {arXiv.org perpetual, non-exclusive license}
}

@article{niche,
  doi = {10.15761/hmo.1000147},
  url = {https://doi.org/10.15761/hmo.1000147},
  year = {2017},
  publisher = {Open Access Text Pvt,  Ltd.},
  volume = {2},
  number = {6},
  author = {Annovazzi Laura and Mellai Marta and Bisogno Ilaria and Spatola Alessandra and Bovio Enrica and Cassoni Paola and Casalone Cristina and Schiffer Davide},
  title = {Perivascular niches as points of the utmost expression of tumor microenvironment},
  journal = {Hematology {\&} Medical Oncology}
}

@article{bull_spatial_stats,
	title = {Combining multiple spatial statistics enhances the description of immune cell localisation within tumours},
	volume = {10},
	issn = {2045-2322},
	url = {https://doi.org/10.1038/s41598-020-75180-9},
	doi = {10.1038/s41598-020-75180-9},
	number = {1},
	journal = {Scientific Reports},
	author = {Bull, Joshua A. and Macklin, Philip S. and Quaiser, Tom and Braun, Franziska and Waters, Sarah L. and Pugh, Chris W. and Byrne, Helen M.},
	month = oct,
	year = {2020},
	pages = {18624},
}

@inproceedings{prostate,
author = {Peter Lawson and Jordan Schupbach and Brittany Terese Fasy and John W. Sheppard},
title = {{Persistent homology for the automatic classification of prostate cancer aggressiveness in histopathology images}},
volume = {10956},
booktitle = {Medical Imaging 2019: Digital Pathology},
editor = {John E. Tomaszewski and Aaron D. Ward},
organization = {International Society for Optics and Photonics},
publisher = {SPIE},
pages = {109560G},
year = {2019},
doi = {10.1117/12.2513137},
URL = {https://doi.org/10.1117/12.2513137}
}

@article{amezquita2020,
author = {Amézquita, Erik J. and Quigley, Michelle Y. and Ophelders, Tim and Munch, Elizabeth and Chitwood, Daniel H.},
title = {The shape of things to come: Topological data analysis and biology, from molecules to organisms},
journal = {Developmental Dynamics},
volume = {249},
number = {7},
pages = {816-833},
doi = {10.1002/dvdy.175},
url = {https://anatomypubs.onlinelibrary.wiley.com/doi/abs/10.1002/dvdy.175},
year = {2020}
}

@book{rabadan_blumberg, place={Cambridge}, 
title={Topological Data Analysis for Genomics and Evolution: Topology in Biology},
DOI={10.1017/9781316671665}, 
publisher={Cambridge University Press}, 
author={Rabadan, Raul and Blumberg, Andrew J.},
year={2019}}

@article{tymochko2020,
  title={Using zigzag persistent homology to detect Hopf bifurcations in dynamical systems},
  author={Tymochko, Sarah and Munch, Elizabeth and Khasawneh, Firas A},
  journal={Algorithms},
  volume={13},
  number={11},
  pages={278},
  year={2020},
  publisher={MDPI},
  url = {https://www.mdpi.com/1999-4893/13/11/278},
  doi = {10.3390/a13110278}
}

@inproceedings{corcoran2016,
author = {Corcoran, Padraig and Jones, Christopher B.},
title = {Spatio-Temporal Modeling of the Topology of Swarm Behavior with Persistence Landscapes},
year = {2016},
isbn = {9781450345897},
publisher = {Association for Computing Machinery},
url = {https://doi.org/10.1145/2996913.2996949},
doi = {10.1145/2996913.2996949},
booktitle = {Proceedings of the 24th ACM SIGSPATIAL International Conference on Advances in Geographic Information Systems},
articleno = {65},
numpages = {4},
series = {SIGSPACIAL '16}
}

@article{mata_zigzag,
title = {Zigzag persistent homology for processing neuronal images},
volume = {62},
issn = {0167-8655},
	url = {https://www.sciencedirect.com/science/article/pii/S0167865515001555},
	doi = {10.1016/j.patrec.2015.05.010},
	journal = {Pattern Recognition Letters},
	author = {Mata, Gadea and Morales, Miguel and Romero, Ana and Rubio, Julio},
	year = {2015},
	pages = {55--60},
}

@Article{epithelial,
author ="Bhaskar, Dhananjay and Zhang, William Y. and Wong, Ian Y.",
title  ="Topological data analysis of collective and individual epithelial cells using persistent homology of loops",
journal  ="Soft Matter",
year  ="2021",
volume  ="17",
issue  ="17",
pages  ="4653-4664",
publisher  ="The Royal Society of Chemistry",
doi  ="10.1039/D1SM00072A",
url  ="http://dx.doi.org/10.1039/D1SM00072A"}

@misc{Dhananjay_heterogeneous,
  doi = {10.48550/ARXIV.2212.14113},
  
  url = {https://arxiv.org/abs/2212.14113},
  
  author = {Bhaskar, Dhananjay and Zhang, William Y. and Volkening, Alexandria and Sandstede, Björn and Wong, Ian Y.},
  
  title = {Topological Data Analysis of Spatial Patterning in Heterogeneous Cell Populations: I. Clustering and Sorting with Varying Cell-Cell Adhesion},
  
  publisher = {arXiv},
  
  year = {2022},
  
  copyright = {Creative Commons Attribution 4.0 International}
}

@Article{neuroblastoma,
AUTHOR = {Bussola, Nicole and Papa, Bruno and Melaiu, Ombretta and Castellano, Aurora and Fruci, Doriana and Jurman, Giuseppe},
TITLE = {Quantification of the Immune Content in Neuroblastoma: Deep Learning and Topological Data Analysis in Digital Pathology},
JOURNAL = {International Journal of Molecular Sciences},
VOLUME = {22},
YEAR = {2021},
NUMBER = {16},
ARTICLE-NUMBER = {8804},
URL = {https://www.mdpi.com/1422-0067/22/16/8804},
PubMedID = {34445517},
ISSN = {1422-0067},
DOI = {10.3390/ijms22168804}
}

@article{lunghistology,
	title = {Topological data analysis of thoracic radiographic images shows improved radiomics-based lung tumor histology prediction},
	volume = {4},
	issn = {2666-3899},
	url = {https://www.sciencedirect.com/science/article/pii/S2666389922002975},
	doi = {10.1016/j.patter.2022.100657},
	language = {en},
	number = {1},
	urldate = {2023-03-09},
	journal = {Patterns},
	author = {Vandaele, Robin and Mukherjee, Pritam and Selby, Heather Marie and Shah, Rajesh Pravin and Gevaert, Olivier},
	month = jan,
	year = {2023},
	pages = {100657},
}

@article{fMRI-vineyard,
title = {Topological persistence vineyard for dynamic functional brain connectivity during resting and gaming stages},
journal = {Journal of Neuroscience Methods},
volume = {267},
pages = {1-13},
year = {2016},
issn = {0165-0270},
doi = {10.1016/j.jneumeth.2016.04.001},
url = {https://www.sciencedirect.com/science/article/pii/S0165027016300395},
author = {Jaejun Yoo and Eun Young Kim and Yong Min Ahn and Jong Chul Ye}}

@article{flamelets,
    author = {Tullia Padellini and Pierpaolo Brutti},
    title = {Persistence Flamelets: Topological Invariants for Scale Spaces},
    journal = {Journal of Computational and Graphical Statistics},
    volume = {32},
    number = {2},
    pages = {671-683},
    year  = {2023},
    publisher = {Taylor & Francis},
    doi = {10.1080/10618600.2022.2074427},
    URL = {https://doi.org/10.1080/10618600.2022.2074427}
}

\end{document}